\documentclass[12pt]{article}
\pdfoutput=1
\usepackage{amssymb,amsmath,graphicx}

\makeatletter

\renewcommand\section{\@startsection {section}{1}{\z@}%
                                 {-3.5ex \@plus -1ex \@minus -.2ex}
                                   {2.3ex \@plus.2ex}%
                                   {\normalfont\large\bfseries}}
\renewcommand\subsection{\@startsection{subsection}{2}{\z@}%
                                   {-3.25ex\@plus -1ex \@minus -.2ex}%
                                     {1.5ex \@plus .2ex}%
                                     {\normalfont\bfseries}}
\renewcommand\subsubsection{\@startsection{subsubsection}{3}{\z@}%
                                   {-3.25ex\@plus -1ex \@minus -.2ex}%
                                     {1.5ex \@plus .2ex}%
                                     {\normalfont\itshape}}
\makeatother

\newcommand{\Letter}{
    \setlength{\textwidth}{7in}
    \setlength{\textheight}{9.3in}
    \hoffset=-0.75in
    \voffset=-1.15in }

\Letter

\newcommand{\non}{\nonumber \\}

\newcommand{\alp}{\alpha}     \newcommand{\bet}{\beta}
\newcommand{\gam}{\gamma}     \newcommand{\del}{\delta}
\newcommand{\eps}{\epsilon}

\newcommand{\kap}{\kappa}

   \newcommand{\ome}{\omega}

\newcommand{\cA}{{\cal A}}

    \newcommand{\cJ}{{\cal J}}
    
    \newcommand{\cN}{{\cal N}}

    \newcommand{\cT}{{\cal T}}

\newcommand{\pa}{\partial}

\renewcommand{\dag}{^{\dagger}}

\newcommand{\rar}{\rightarrow}

\newcommand{\one}{1\!\!1}

\newcommand{\tr}{\mbox{Tr}}

\newcommand{\hlf}{\frac{1}{2}}

\newcommand{\gsim}{ \lower .75ex \hbox{$\sim$} \llap{\raise .27ex \hbox{$>$}} }
\newcommand{\lsim}{ \lower .75ex \hbox{$\sim$} \llap{\raise .27ex \hbox{$<$}} }

\def\slT{\,\slash\!\!\!\! \cT}
\def\slD{\slash\!\!\!\!D}

\begin{document}
\thispagestyle{empty}
\begin{flushright}
\parbox[t]{3.5in}{
{\bf ***~\jobname.tex~***}}
\end{flushright}

\vspace*{0.5in}

\begin{center}
{\large \bf 
Constructing the AdS dual of a Fermi liquid:\\
AdS Black holes with Dirac hair}\\

\vspace*{0.5in} {Mihailo~\v{C}ubrovi\'{c},
Jan~Zaanen, Koenraad~Schalm
}\\[.3in]
{\em
     ${}^1$ Institute Lorentz for Theoretical Physics, Leiden University\\
P.O. Box 9506, Leiden 2300RA, The Netherlands \\[.3in]
}
\end{center}

\begin{center}
{\bf
Abstract}
\end{center}
\noindent We provide new evidence that the holographic dual to a
strongly coupled charged Fermi liquid has a non-zero fermion
density in the bulk. We show that the pole-strength  of the stable
quasiparticle characterizing the Fermi surface is encoded in the
spatially averaged AdS probability density of a single
normalizable fermion wavefunction in AdS. Recalling Migdal's
theorem which relates the pole strength to the Fermi-Dirac
characteristic discontinuity in the number density at $\ome_F$, we
conclude that the AdS dual of a Fermi liquid is described by
occupied on-shell fermionic modes in AdS. Encoding the occupied
levels in the total probability density of the fermion field
directly, we show that an AdS Reissner-Nordstr\"{o}m black hole in
a theory with charged fermions has a critical temperature, at
which the system undergoes a first-order transition to a black
hole with a non-vanishing profile for the bulk fermion field.
Thermodynamics and spectral analysis confirm that the solution
with non-zero AdS fermion-profile is the preferred ground state at
low temperatures. \vfill

\hrulefill\hspace*{4in}

{\footnotesize
Email addresses: \parbox[t]{5.5in}{cubrovic, jan, kschalm@lorentz.leidenuniv.nl.
}
}

\newpage
\section{Introduction}

Fermionic quantum criticality is thought to be an essential
ingredient in the full theory of high $T_c$ superconductivity
\cite{SFL,SZ}.  The cleanest experimental examples of quantum
criticality occur in heavy-fermion systems rather than high $T_c$
cuprates, but the experimental measurements in heavy fermions
raise equally confounding theoretical puzzles \cite{RV}. Most
tellingly, the resistivity scales linearly with the temperature
from the onset of superconductivity up to the crystal melting
temperature \cite{linear} and this linear scaling is in conflict
with single correlation length scaling at criticality
\cite{PhillipsChamon}. The failure of standard perturbative
theoretical methods to describe such behavior is thought to
indicate that the underlying quantum critical system is strongly
coupled \cite{JZScience,Luttinger}.

The combination of strong coupling and scale-invariant critical
dynamics makes these systems an ideal arena for the application of
the AdS/CFT correspondence: the well-established relation between
strongly coupled conformal field theories (CFT) and gravitational
theories in anti-de Sitter (AdS) spacetimes. An AdS/CFT
computation of single-fermion spectral functions ---  which are
directly experimentally accessible via Angle-Resolved
Photoemission Spectroscopy \cite{arpes1,arpes2,arpes3} --- bears
out this promise of addressing fermionic quantum criticality
\cite{Liu:2009dm,ours,Faulkner:2009wj,strangemet} (see also
 \cite{Lee:2008xf,Rey:2009ag}).
The AdS/CFT single fermion spectral function exhibits distinct
sharp quasiparticle peaks, associated with the formation of a
Fermi surface, emerging from a scale-free state. The fermion
liquid which this Fermi surface captures is generically singular:
it has either a non-linear dispersion or non-quadratic pole
strength \cite{Liu:2009dm,Faulkner:2009wj}.
The precise
details depend on the parameters of the AdS model.

From the AdS gravity perspective, peaks with linear dispersion
correspond to the existence of a stable charged fermionic
quasinormal mode in the spectrum of a charged AdS black hole. The
existence of a stable charged bosonic quasinormal mode is known to
signal the onset of an instability towards a new ground state with
a pervading Bose condensate extending from the charged black hole
horizon to the boundary of AdS. The dual CFT description of this
charged condensate is spontaneous symmetry breaking as in a
superfluid and a conventional superconductor
\cite{Gubser:2008px,Hartnoll:2008vx,Hartnoll:2008kx,Hartnoll:2009sz}.
For fermionic systems empirically the equivalent robust $T=0$
groundstate is the Landau Fermi Liquid --- the quantum groundstate
of a system with a finite number of fermions. The existence of a
stable fermionic quasinormal mode suggests that an AdS dual of a
finite fermion density state exists.

We construct here the AdS/CFT rules for CFTs with a finite fermion
density. The essential ingredient will be the Migdal's theorem, which
relates the characteristic jump in fermion occupation number at
the energy $\ome_F$ of the highest occupied state to the pole
strength of the quasiparticle. The latter we know from the
spectral function analysis and its AdS formulation is therefore
known. Using this, we show that the fermion number discontinuity
is encoded in the {\em probability density} of the normalizable
wavefunction of the dual AdS fermion field.

These AdS/CFT rules prove that the AdS dual of a Fermi liquid is
given by a system with occupied fermionic states in the bulk. The
Fermi liquid is clearly not a scale invariant state, but any such
states will have energy, momentum/pressure and charge and will
change the interior geometry from AdS to something else. Which
particular (set of) state(s) is the right one, it does not yet
tell us, as this conclusion relies only on the asymptotic behavior
of fermion fields near the AdS boundary. Here we shall take the
simplest such state: a single
fermion.\footnote{These solutions are therefore the AdS extensions
of
\cite{Finster:2002vk,Finster:2000vy,Finster:2000ps,Finster:1999mn}.}
Constructing the associated backreacted asymptotically AdS
solution, we find that it is already good enough to solve several
problems of principle:
\begin{itemize}
\item Fermionic quantum critical systems should undergo a phase
transition to a Fermi liquid at low temperatures, except when one
is directly above the QCP. Likewise we find that the dual of the
quantum critical state, a charged AdS black hole in the presence
of charged fermionic modes, has a critical temperature below which
fermionic Dirac ``hair'' forms. The derivative of the free energy
has the characteristic discontinuity of a first order transition.
This has to be the case:
A fermionic quasinormal mode can never cause a linear instability
indicative of a continuous phase transition. In the language of
spectral functions, the pole of the retarded Green's function can
never cross to the upper-half plane \cite{Faulkner:2009wj}.\footnote{Ref. \cite{Cai:2010tr} argues that the instability can be second order.} The
absence of a perturbative instability between this conjectured
Dirac ''black hole hair'' solution and the ``bald'' charged AdS
black hole can be explained if the transition is a first order
gas-liquid transition. The existence of first order transition
follows from a thermodynamic analysis of the free energy rather
than a spectral analysis of small fluctuations.

\item This solution with finite fermion number is the preferred
ground state at low temperatures. The bare charged AdS black hole
in a theory with charged fermions is therefore a false vacuum.
Confusing a false vacuum with the true ground state can lead to
anomalous results. Indeed the finite temperature behavior of
fermion spectral functions in AdS Reissner-Nordstr\"{o}m,
exhibited in the combination of the results of
\cite{Liu:2009dm,Faulkner:2009wj} and \cite{ours}, shows strange
behavior. The former \cite{Liu:2009dm,Faulkner:2009wj} found sharp
quasiparticle peaks at a frequency $\ome_F=0$ in natural AdS
units, whereas the latter \cite{ours} found sharp quasiparticle
peaks at finite Fermi energy $\ome_F\neq 0$. As we will show, both
peaks in fact describe the same physics: the $\ome_F\neq 0$ peak
is a finite temperature manifestation of (one of the) $\ome=0$
peaks in \cite{Faulkner:2009wj}. Its shift in location at finite
temperature is explained by the existence of the nearby true
finite fermion density ground state, separated by a potential
barrier from the AdS Reissner-Nordstr\"{o}m solution.

\item The charged AdS-black hole solution corresponds to a CFT
system in a state with large ground state entropy. This is the
area of the extremal black-hole horizon at $T=0$. Systems with
large ground-state entropy are notoriously unstable to collapse to
a low-entropy state, usually by spontaneous symmetry breaking. In
a fermionic system it should be the collapse to the Fermi liquid.
The final state will generically be a geometry that asymptotes to
Lifschitz type, i.e. the background breaks Lorentz-invariance and
has a double-pole horizon with vanishing area, as expounded in
\cite{polchinski}. Although the solution we construct here only
considers the backreaction on the electrostatic potential, we show
that the gravitational energy density diverges at the horizon in a
similar way as other systems that are known to gravitationally
backreact to a Lifshitz solution. The fully backreacted geometry
includes important separate physical aspects --- it is relevant to
the stability of the Fermi liquid --- and will be considered in a
companion article.
\end{itemize}

The Dirac hair solution thus captures the physics one expects of
the dual of a Fermi liquid. We have based its construction on a
derived set of AdS/CFT rules to describe systems at finite fermion
density. Qualitatively the result is as expected: that one also
needs occupied fermionic states in the bulk. Knowing this, another
simple candidate is the dual of the backreacted AdS-Fermi-gas
\cite{polchinski}/electron star \cite{electronstar} which appeared
during the course of this work.\footnote{See also
\cite{verlinde,verlinde2}. An alternative approach to back-reacting fermions is \cite{Jatkar}.}
The difference between the two approaches are the assumptions used to reduce the interacting Fermi system to a tractable solution.
As explained in the recent article \cite{spec},
the Fermi-gas and the single Dirac field are the two ``local'' approximations to the generic non-local multiple fermion system in the bulk, in very different regimes of applicability. The electron-star/Fermi-gas is considered in the Thomas-Fermi limit where the microscopic charge of the constituent fermions is sent to zero keeping the overall charge fixed, whereas the single Dirac field clearly is the 'limit' where the microscopic charge equals the total charge in the system. This is directly evident in the spectral functions of both systems. The results presented here show that each pole in the CFT spectral function corresponds to a unique occupied Fermi state in the bulk; the electron star spectra show a parametrically large number of poles \cite{sean,hong,spec}, whereas the Dirac hair state has a single quasiparticle pole by construction.
The AdS-Dirac-hair black hole derived here therefore has the benefit of a direct connection with a unique Fermi liquid state in the CFT. This is in fact the starting point of our derivation.

In the larger context, the existence of both the Dirac hair and backreacted Fermi
gas solution is not a surprise. It is a manifestation of {\em
universal} physics in the presence of charged AdS black holes. The
results here, and those of
\cite{Liu:2009dm,Faulkner:2009wj,polchinski,electronstar},
together with the by now extensive literature on holographic
superconductors, i.e. Bose condensates, show that at sufficiently
low temperature in units of the black-hole charge, the electric
field stretching to AdS-infinity causes a spontaneous discharge of
the bulk vacuum outside of the horizon into the charged fields of
the theory --- whatever their nature. The positively charged
excitations are repelled by the black hole, but cannot escape to
infinity in AdS and they form a charge cloud hovering over the
horizon. The negatively charged excitations fall into the
black-hole and neutralize the charge, until one is left with an
uncharged black hole with a condensate at finite $T$ or a pure
asymptotically AdS condensate solution at $T=0$. As
\cite{polchinski,electronstar} and we show, the statistics of the
charged particle do not matter for this condensate formation,
except in the way it forms: bosons superradiate and fermions
nucleate. The dual CFT perspective of this process is ``entropy
collapse''. The final state therefore has negligible ground state
entropy and is stable. The study of charged black holes in AdS/CFT is
therefore
 a novel way to understand the stability of charged
interacting matter which holds much promise.



\section{From Green's function to AdS/CFT rules for a Fermi Liquid}
\label{sec:hologr-migd-theor}

We wish to show how a solution with finite fermion number --- a Fermi liquid --- is encoded in AdS.  The exact connection and derivation will require a review of what we have learned of Dirac field dynamics in AdS/CFT through Green's functions analysis. 
The defining signature of a Fermi liquid is a quasi-particle pole in
the (retarded) fermion propagator,
\begin{equation}
\label{zprop}G_R=\frac{Z}{\omega-\mu_R-v_F(k-k_F)}+ \mbox{regular}
\end{equation}
Phenomenologically, a non-zero residue at the pole, $Z$, also known
as the pole strength, is the indicator of a Fermi liquid state.
Migdal famously related the pole strength to the occupation number
discontinuity at the pole ($\ome=0$):
\begin{eqnarray}
  \label{eq:17}
  Z = \lim_{\eps\rar 0} \left[n_F(\ome-\eps)-n_F(\ome+\eps)\right]
\end{eqnarray}
where $$n_F(\ome)=\int d^2 k f_{FD}\left(\frac{\ome}{T}\right)
{\rm Im} G_R(\ome, k).$$
where $f_{FD}$ is the Fermi-Dirac distribution function. Vice versa, a Fermi liquid with a Fermi-Dirac jump in occupation
number at the Fermi energy $\ome_F=0$ has a low-lying
quasiparticle excitation.
Using our knowledge of fermionic spectral functions in AdS/CFT we
shall first relate the pole-strength $Z$ to known AdS quantities.
Then, using Migdal's relation, the dual of a Fermi liquid is
characterized by an asymptotically AdS solution with non-zero
value for these very objects.

The Green's functions derived in AdS/CFT are those of charged fermionic operators with scaling dimension $\Delta$,
dual to an AdS Dirac field with mass
$m=\Delta-\frac{d}{2}$. We shall focus on $d=2+1$ dimensional
CFTs. In its gravitational description, this Dirac field is
minimally coupled to $3+1$ dimensional gravity and
electromagnetism with the action
\begin{eqnarray}
  \label{eq:2a}
  S= \int d^4 x \sqrt{-g} \left[\frac{1}{2\kap^2}\left( R + \frac{6}{L^2} \right) -\frac{1}{4} F_{MN}^2 - \bar{\Psi}(\slash\!\!\!\!D+m)\Psi \right]~.
\end{eqnarray}
For zero background fermions, $\Psi=0$, a spherically symmetric solution is a
charged AdS${}_4$ black-hole background
\begin{eqnarray}
\label{metric}ds^2&=&\frac{L^2\alpha^2}{z^2}\left(-f(z)dt^2+dx^2+dy^2\right)+\frac{L^2}{z^2}\frac{dz^2}{f(z)}~,\non
f(z)&=&(1-z)(1+z+z^2-q^2z^3)~,\non A_0^{(bg)}&=&2q\alp(z-1)~.
\end{eqnarray}
Here $A_0^{(bg)}$ is the time-component of the
$U(1)$-vector-potential, $L$ is the AdS radius and the temperature
and chemical potential of the black hole equal
\begin{eqnarray}
  \label{eq:1}
  T = \frac{\alpha}{4\pi}(3-q^2)~,~~\mu_0=-2q\alp,
\end{eqnarray}
where $q$ is the black hole charge.

To compute the Green's functions we need to solve the Dirac equation in the background of this charged black hole:
\begin{eqnarray}
  \label{eq:2}
e_A^M\Gamma^A(D_M+i{\rm e}gA_M)\Psi+m\Psi=0~,
\end{eqnarray}
where the vielbein $e_A^M$, covariant derivative $D_M$ and connection $A_M$
correspond to the fixed charged AdS black-hole metric and electrostatic potential
(\ref{metric}) and $g$ is the fermion charge. 
Denoting $A_0=\Phi$ and taking the standard
AdS-fermion projection onto $\Psi_{\pm}=\hlf(1\pm\Gamma^{Z})\Psi$,
the Dirac equation reduces to
\begin{eqnarray}
\label{de-org}
\left(\partial_z+\mathcal{A}_\pm\right)\Psi_\pm&=&\mp\slT\Psi_\mp
\end{eqnarray}
with
\begin{eqnarray}
\label{not1}\mathcal{A}_\pm&=&-\frac{1}{2z}\left(3-\frac{zf^\prime}{2f}\right)\pm\frac{mL}{z\sqrt{f}}~, \non
\label{not2}\slT&=&\frac{i(-\ome +g\Phi)}{\alpha f}\gamma^0+\frac{i}{\alp\sqrt{f}}k_i\gamma^i~.
\end{eqnarray}
Here $\gamma^{\mu}$ are the 2+1-dimensional Dirac matrices, obtained
after decomposing the 3+1 dimensional $\Gamma^{\mu}$-matrices.

Explicitly the Green's function is extracted from the behavior of
the solution to the Dirac equation at the AdS-boundary. The boundary
behavior of the bulk fermions is
\begin{eqnarray}
  \label{eq:4}
  \Psi_+(\ome, k;z) &=& A_+z^{\frac{3}{2}-m} + B_+z^{\frac{5}{2}+m} +\ldots,\non
\Psi_-(\ome,k;z) &=& A_-z^{\frac{5}{2}-m} + B_-z^{\frac{3}{2}+m} +\ldots,
\end{eqnarray}
where $A_\pm(\ome,k),~B_\pm(\ome,k)$  are not all independent but related by the
Dirac equation at the boundary
\begin{eqnarray}
\label{eq:5}
A_-= -\frac{i\mu}{(2m-1)}\gamma^0A_+~,~~B_+= -\frac{i\mu}{(2m+1)}\gamma^0B_-~.
\end{eqnarray}
The CFT Green's function then equals
\cite{ours,Iqbal:2009fd,Liu:2009dm}
\begin{eqnarray}
  \label{eq:14}
  G_R = \lim_{z\rar 0}  z^{-2m}\frac{\Psi_-(z)}{\Psi_+(z)} - {\rm singular} =  \frac{B_-}{A_+}~.
\end{eqnarray}
In other words, $B_-$ is the CFT response to the (infinitesimal) source $A_+$.
Since in the Green's function the fermion is a fluctuation, the
functions $\Psi_{\pm}(z)$ are now probe solutions to the Dirac
equation in a fixed gravitational and electrostatic background
(for ease of presentation we are considering $\Psi_{\pm}(z)$ as
numbers instead of two-component vectors). The boundary conditions
at the horizon/AdS interior determine which Green's function one
considers, e.g. infalling horizon boundary conditions yield the
retarded Green's function. For non-zero chemical potential this
fermionic Green's function can have a pole signalling the presence
of a Fermi surface. This pole occurs precisely for a
\mbox{(quasi-)normalizable} mode, i.e. a specific energy $\ome_F$
and momentum $k_F$ where the external source $A_+(\ome,k)$
vanishes (for infalling boundary conditions at the horizon). 

Knowing that the energy of the quasinormal mode is always $\ome_F=0$ \cite{Liu:2009dm} and following \cite{Faulkner:2009wj},
we expand $G_R$ around $\ome=0$ as:
\begin{equation}
\label{match}G_R(\omega)= \frac{B^{(0)}+\ome B^{(1)}+\ldots}{A_{+}^{(0)}+\omega A_{+}^{(1)}+\ldots}.
\end{equation}
A crucial point is that in this expansion we are assuming that the
pole will correspond to a stable quasiparticle, i.e. there are no
fractional powers of $\ome$ less than unity in the expansion
around $\ome_F=0$ \cite{Faulkner:2009wj}. Fermions in AdS/CFT are
of course famous for allowing more general pole-structures
corresponding to Fermi-surfaces without stable quasiparticles \cite{Faulkner:2009wj}, but
those Green's functions are not of the type \eqref{zprop} and we
shall therefore not consider them here. The specific Fermi
momentum $k_F$ associated with the Fermi surface is the momentum
value for which the first $\ome$-independent term in the
denominator vanishes $A_{+}^{(0)}(k_F)=0$ --- for this value of
$k=k_F$ the presence of a pole in the Green's functions at
$\ome=0$ is manifest. Writing $A_+^{(0)} = a_+(k-k_F) + \ldots$
and comparing with the standard quasi-particle propagator,
\begin{equation}
\label{zprop2}G_R=\frac{Z}{\omega-\mu_R-v_F(k-k_F)}+ \mbox{regular}
\end{equation}
we read off that the pole-strength equals $$Z=B^{(0)}_-(k_F)/A_+^{(1)}(k_F).$$

We thus see that a non-zero pole-strength is ensured by a non-zero
value of \mbox{$B_-(\ome=0,k=k_F)$} --- the ``response'' {\em
without} corresponding source as $A^{(0)}(k_F)\equiv 0$.
Quantatively the pole-strength also depends on the value of $A_+^{(1)}(k_F)\equiv \pa_\ome A_+(k_F)|_{\ome=0}$, which is always finite. This is not a truly independent parameter, however. The size of the pole-strength has  only a relative meaning w.r.t. to the integrated spectral density. 
This normalization of the pole strength is a global parameter
rather than an AdS boundary issue. We now show this by proving
that $A_+^{(1)}(k_F)$ is inversely proportional to
$B_-^{(0)}(k_F)$ and hence $Z$ is completely set by
$B_-^{(0)}(k_F)$, i.e. $Z \sim |B_-^{(0)}(k_F)|^2$. Consider a
transform $\widetilde{W}({\Psi}_{+,A},\Psi_{+,B})$ of the
Wronskian
$W({\Psi}_{+,A},\Psi_{+,B})={\Psi}_{+,A}\partial_z\Psi_{+,B}-(\pa_z{\Psi}_{+,A})\Psi_{+,B}$
for two solutions to the second order equivalent of the Dirac
equation for the field $\Psi_+$
\begin{eqnarray}
\left(\pa_z^2  + P(z)\pa_z + Q_+(z)\right)\Psi_+=0
\label{eq:46}
\end{eqnarray}
that is conserved (detailed expressions for $P(z)$ and $Q_+(z)$ are given in eq. (\ref{eq:15})):
\begin{eqnarray}
\widetilde{W}(\Psi_{+,A}(z),\Psi_{+,B}(z),z;z_0)=\exp\left(\int^z_{z_0} P(z)\right)W(\Psi_{+,A}(z),\Psi_{+,B}(z))~,~~~~\pa_z\widetilde{W}=0.\label{eq:47}
\end{eqnarray}
Here $z_0^{-1}$ is the infinitesimal distance away from the
boundary at $z=0$ which is equivalent to the $UV$-cutoff in the
CFT. Setting $k=k_F$ and choosing for $\Psi_{+,A}= A_+
z^{3/2-m}\sum_{n=0}^{\infty}a_nz^n$ and
$\Psi_{+,B}=B_+z^{5/2+m}\sum_{n=0}^{\infty}b_nz^nr$ the real
solutions which asymptote to solutions with $B_+(\ome,k_F)=0$ and
$A_+(\ome,k_F)=0$ respectively, but for a value of $\ome$
infinitesimally away from $\ome_F=0$, we can evaluate
 $\widetilde{W}$ at the boundary to find,\footnote{$P(z)=-3/z+\ldots$ near $z=0$}
\begin{eqnarray}
  \label{eq:48}
 \widetilde{W}=z_0^3(1+2m){A}_+B_+ 
= \mu z_0^3 A_+B_-
\end{eqnarray}
The last step follows from the constraint \eqref{eq:5} where the
reduction from two-component spinors to functions means that
$\gamma^0$ is replaced by one of its eigenvalues $\pm i$. Taking
the derivative of $\widetilde{W}$ at $\ome=0$ for $k=k_F$ and
expanding $A_+(\ome,k_F)$ and $B_-(\ome,k_F)$ as in \eqref{match},
we can solve for $A_+^{(1)}(k_F)$ in terms of $B_-^{(0)}(k_F)$ and
arrive at the expression for the pole strength $Z$ in terms of
$|B_-^{(0)}(k_F)|^2$:
\begin{equation}
\label{zval}Z=\frac{\mu z_0^{3}}{\partial_\omega\widetilde{W}|_{\ome=0,k=k_F}}|B_-^{(0)}(k_F)|^2~.
\end{equation}
Because $\partial_\omega\widetilde{W}$, as
$\widetilde{W}$, is a number that is
 independent of $z$, this expression emphasizes that it is truly the nonvanishing subleading term $B_-^{(0)}(\ome_F,k_F)$ which sets the pole strength, up to a normalization $\partial_\omega\widetilde{W}$ which is set by the fully integrated spectral density. This integration is always UV-cut-off dependent and the explicit $z_0$ dependence should therefore not surprise us.\footnote{Using that $\widetilde{W}$ is conserved, one can e.g. compute it at the horizon. There each solution $\Psi_{+,A}(\ome,k_F;z)$, $\Psi_{+,B}(\ome,k_F;z)$  is a linear combination of the infalling and outgoing solution
\begin{eqnarray}
\label{horwron}
\Psi_{+,A}(z)&=&
 \bar{\alp}\left(1-z\right)^{-1/4+\imath{\omega}/4\pi T}+
{\alp}\left(1-z\right)^{-1/4-\imath{\omega}/4\pi T}+\ldots \non
 \Psi_{+,B}(z)&=&
 \bar{\bet}\left(1-z\right)^{-1/4+\imath{\omega}/4\pi T}+{\bet}\left(1-z\right)^{-1/4-\imath{\omega}/4\pi T}+\ldots
\end{eqnarray}
yielding a value of 
$\pa_{\ome}\widetilde{W}$ equal to ({$P(z)=1/2(1-z)+\ldots$ near $z=1$})
\begin{equation}
\label{horwron2}
\pa_{\ome}\widetilde{W}=\frac{i}{2\pi T}\cN(z_0)(\bar{\alp}\bet-\bar{\bet}{\alp})
\end{equation}
with $\cN(z_0) = \exp{\int_{z_0}^{z}dz\left[P(z)-\frac{1}{2(1-z)}\right]}$.
}
We should note that, unlike perturbative Fermi liquid theory, $Z$
is a dimensionful quantity of mass dimension $2m+1=2\Delta-2$,
which illustrates more directly its scaling dependence on the
UV-energy scale $z_0$. At the same time $Z$ is real, as it can be
shown that both $\pa_\ome \widetilde{W}|_{\ome=0,k=k_F} = \mu
z_0^3 A_+^{(1)}B_-^{(0)}$ and $B_-^{(0)}$ are real
\cite{Faulkner:2009wj}.

\subsection{The AdS dual of a stable Fermi Liquid: Applying Migdal's relation holographically}
\label{sec:dynam-dirac-field}


We have thus seen that a solution with nonzero $B_-(\ome_F,k_F)$
whose corresponding external source vanishes (by definition of
$\ome_F,~k_F$), is related to the presence of a quasiparticle pole
in the CFT. Through Migdal's theorem its pole strength is related
to the presence of a discontinuity of the occupation number, and
this discontinuity is normally taken as the characteristic
signature of the presence of a Fermi Liquid. Qualitatively we can
already infer that an AdS gravity solution with non-vanishing
$B_-(\ome_F,k_F)$ corresponds to a Fermi Liquid in the CFT. We
thus seek solutions to the Dirac equation 
with
vanishing external source $A_+$ but non-vanishing response $B_-$
coupled to electromagnetism (and gravity).  The
construction of the AdS black hole solution with a finite single
fermion wavefunction is thus analogous to the construction of a
holographic superconductor \cite{Hartnoll:2008vx} with the role of
the scalar field now taken by a Dirac field of mass $m$.

This route is complicated, however, by the spinor representation of the
Dirac fields, and the related fermion doubling in AdS. Moreover, relativistically the fermion Green's function is a matrix and
the pole strength $Z$ appears in the time-component of the vector projection Tr$i\gam^iG$. As
we take this and the equivalent jump in occupation number to be
the signifying characteristic of a Fermi liquid state in the CFT,
it would be much more direct if we can derive an AdS radial
evolution equation for the vector-projected Green's function and hence the
occupation number discontinuity directly. From the AdS
perspective is also more convenient to work with bilinears such as Green's functions, since
the Dirac fields always couple pairwise to bosonic fields.

To do so, we start again with the two decoupled second order equations equivalent to the Dirac equation \eqref{de-org}
\begin{eqnarray}
  \label{eq:14b}
\left(\pa_z^2+P(z)\pa_z +Q_{\pm}(z)\right)\Psi_{\pm}=0
\end{eqnarray}
with
\begin{eqnarray}
\label{eq:15}
&&P(z) = (\cA_-+\cA_+)-[\pa_z,\slT]\frac{\slT}{T^2}~, \non
&&Q_{\pm}(z) = \cA_-\cA_++(\pa_z\cA_{\pm})-[\pa_z,\slT]\frac{\slT}{T^2}\cA_{\pm}+T^2~.
\end{eqnarray}
Note that both $P(z)$ and $Q_{\pm}(z)$ are matrices in spinor
space. The general solution to this second order equation ---
{with the behavior at the horizon/interior appropriate for the
Green's function one desires} --- is a matrix-valued function
$(M_{\pm}(z))^{\alp}_{~\bet}$ and the field $\Psi_{\pm}(z)$ equals
$\Psi_{\pm}(z) = M_{\pm}(z)\Psi_{\pm}^{(hor)}$. Due to the first
order nature of the Dirac equation the horizon values
$\Psi_{\pm}^{(hor)}$ are not independent but related by a
$z$-independent matrix $S\Psi_+^{(hor)}=\Psi_-^{(hor)}$, which can
be deduced from the near-horizon behavior of \eqref{eq:5};
specifically $S=\gamma^0$. One then obtains the Green's function
from the on-shell boundary action (see e.g. \cite{contino,ours})
\begin{eqnarray}
  \label{eq:7a}
  S_{bnd} = \oint_{z=z_0} d^dx  \bar{\Psi}_+\Psi_-
\end{eqnarray}
as follows:
Given a boundary source $\zeta_{+}$ for $\Psi_+(z)$, i.e. $\Psi_+(z_0)\equiv \zeta_+$, one concludes that $\Psi^{(hor)}_+=M^{-1}_+(z_0)\zeta_+$ and thus
$\Psi_+(z)=M_+(z)M_+^{-1}(z_0)\zeta_+$, $\Psi_-(z) = M_-(z)SM^{-1}_+(z_0)\zeta_+$.
Substituting these solutions into the action gives
\begin{eqnarray}
  \label{eq:3}
  S_{bnd} =\oint_{z=z_0} d^dx \,\bar{\zeta}_+ M_-(z_0)SM_+^{-1}(z_0)\zeta_+
\end{eqnarray}
The Green's function is  obtained by differentiating w.r.t.
$\bar{\zeta}_+$ and $\zeta_+$ and discarding the conformal factor
$z_0^{2m}$ with $m$ being the AdS mass of the Dirac field (one has to be
careful for $mL>1/2$ with analytic terms \cite{contino})
\begin{eqnarray}
  \label{eq:6c}
  G = \lim_{z_0\rar 0} z_0^{-2m} M_-(z_0)SM_+^{-1}(z_0)~.
\end{eqnarray}

Since $M_{\pm}(z)$ are determined by evolution equations in $z$,
it is clear that the Green's function itself is also determined by
an evolution equation in $z$, i.e. there is some function $G(z)$
which reduces in the limit $z\rar 0$  to $z_0^{2m}G$. One obvious
candidate is the function
\begin{eqnarray}
  \label{eq:8}
  G^{(obv)}(z) = M_-(z)SM_+^{-1}(z)~.
\end{eqnarray}
Using the original Dirac equations one can see that this function
obeys the non-linear evolution equation
\begin{eqnarray}
  \label{eq:9b}
  \pa_z G^{(obv)}(z) = -A_- G^{(obv)}(z) - \slT M_+SM_+^{-1} +A_+ G^{(obv)}(z) + G^{(obv)}(z)\slT G^{(obv)}(z) ~.
\end{eqnarray}
This is the approach used in \cite{Liu:2009dm}, where a specific
choice of momenta is chosen such that $M_+$ commutes with $S$. For
a generic choice of momenta, consistency requires that one also
considers the evolution equation for $M_+(z)SM_+^{-1}(z)$.

There is, however, another candidate for the extension $G(z)$
which is based on the underlying boundary action. Rather than
extending  the kernel $M_-(z_0)M_+^{-1}(z_0)$ of the boundary
action we extend the constituents of the action itself, based on
the individual fermion wavefunctions $\Psi_{\pm}(z) =
M_{\pm}(z)S^{\hlf \mp \hlf }M_{+}^{-1}(z_0)$. We define an
extension of the matrix $G(z)$ including an expansion in the
complete set
$\Gamma^I=\{\one,\gamma^i,\gamma^{ij},\ldots,\gamma^{i_1,i_d}\}$
(with $\gamma^4=i\gamma^0$)
\begin{eqnarray}
  \label{eq:10b}
  {G}^I(z)= \bar{M}_+^{-1}(z_0) \bar{M}_+(z) \Gamma^I M_{-}(z)SM_{+}^{-1}(z_0)~,~~{G}^I(z_0)=\Gamma^I G(z_0)
\end{eqnarray}
where $\bar{M} = i\gamma^0M^{\dagger}i\gamma^0$.  Using again the
original Dirac equations, this function obeys the evolution
equation
\begin{eqnarray}
  \label{eq:11b}
  \pa_z {G}^I(z) = - (\bar{A}_++A_-)G^I(z) -\bar{M}_{+,0}^{-1} \bar{M}_-(z)\bar{\slT}\Gamma^I M_{-}(z)S{M}_{+,0}^{-1} +\bar{M}_{+,0}^{-1} \bar{M}_+(z) \Gamma^I\slT M_{+}(z)S{M}_{+,0}^{-1} \non 
\end{eqnarray}
Recall that $\slT \gamma^{i_1\ldots i_p}=\cT^{[i_1}\gamma^{\ldots
i_p]}+\cT_j\gamma^{ji_1\ldots i_p}$. It is then straightforward to
see that for consistency, we also need to consider the evolution
equations of $$\cJ_{+}^I = \bar{M}_{+,0}^{-1}
\bar{M}_{+}(z)\Gamma^I M_{+}(z)S{M}_{+,0}^{-1} ~,~~~\cJ_-^I =
\bar{M}_{+,0}^{-1} \bar{M}_{-}(z)\Gamma^I
M_{-}(z)S{M}_{+,0}^{-1}$$
 and
$$\bar{G}^I = \bar{M}_{+,0}^{-1} \bar{M}_{-}(z)\Gamma^I M_{+}(z)S{M}_{+,0}^{-1}.$$ They are
\newcommand{\llong}[1]{#1}
\renewcommand{\llong}[1]{}
\begin{eqnarray}
  \label{eq:12b}
  \pa_{z} \cJ_{+}^{i_1\ldots i_p}(z)
\llong{&=& -(\bar{\mathcal{A}}_{+}+\mathcal{A}_{+})\cJ_{+}^{i_1\ldots i_p} - \bar{M}_{+,0}^{-1} \bar{M}_{-}(z)\bar{\slT}\gamma^{i_1\ldots i_p} M_{+}(z)S{M}_{+,0}^{-1} - \bar{M}_{+,0}^{-1} \bar{M}_{+}(z) \gamma^{i_1\ldots i_p} \slT M_{-}(z)S{M}_{+,0}^{-1} \non
}
&=&-2{\rm Re}(\mathcal{A}_{+})\cJ_{+}^{i_1\ldots i_p} - \bar{\cT}^{[i_1}\bar{G}^{i_2\ldots i_p]}(z)-\bar{\cT}_{j}\bar{G}^{ji_1\ldots i_p}(z) - {G}^{[i_1\ldots i_{p-1}}(z){\cT}^{i_p]}-{G}^{i_1\ldots i_pj}(z) {\cT}_{j} \non
\pa_{z} \cJ_{-}^{i_1\ldots i_p}(z)
\llong{&=& -(\bar{A}_{-}+A_{-})\cJ_{-}^{i_1\ldots i_p} + \bar{M}_{+,0}^{-1} \bar{M}_{+}(z)\bar{\slT}\gamma^{i_1\ldots i_p} M_{-}(z)S{M}_{+,0}^{-1} + \bar{M}_{+,0}^{-1} \bar{M}_{-}(z) \gamma^{i_1\ldots i_p} \slT M_{+}(z)S{M}_{+,0}^{-1} \non
}
&=&-2{\rm Re}(\mathcal{A}_{-})\cJ_{-}^{i_1\ldots i_p} + \bar{\cT}^{[i_1}{G}^{i_2\ldots i_p]}(z)+\bar{\cT}_{j}{G}^{ji_1\ldots i_p}(z) + \bar{G}^{[i_1\ldots i_{p-1}}(z){\cT}^{i_p]}+\bar{G}^{i_1\ldots i_pj}(z) {\cT}_{j} \non
\pa_{z} \bar{G}^{i_1\ldots i_p}(z)
\llong{&=& -(\bar{A}_{-}+A_{+})\bar{G}^{i_1\ldots i_p} - \bar{M}_{+,0}^{-1} \bar{M}_{+}(z)\bar{\slT}\gamma^{i_1\ldots i_p} M_{+}(z)S{M}_{+,0}^{-1} - \bar{M}_{+,0}^{-1} \bar{M}_{-}(z) \gamma^{i_1\ldots i_p} \slT M_{+}(z)S{M}_{+,0}^{-1} \non
}
&=&-(\bar{\mathcal{A}}_{-}+\mathcal{A}_{+})\bar{G}^{i_1\ldots i_p} - \bar{\cT}^{[i_1}\cJ_+^{i_2\ldots i_p]}(z)-\bar{\cT}_{j}\cJ_+^{ji_1\ldots i_p}(z) - \cJ_-^{[i_1\ldots i_{p-1}}(z){\cT}^{i_p]}+\cJ_-^{i_1\ldots i_pj}(z) {\cT}_{j} \non
\end{eqnarray}
The significant advantage of these functions
$G^I,~\bar{G}^I,~\cJ_{\pm}^I$ is that the evolution equations are
now linear. This approach may seem overly complicated. However, if
the vector $\cT^i$ happens to only have a single component
nonzero, then the system reduces drastically to the four fields
$\cJ_{\pm}^i,G^{\one}, \bar{G}^{\one}$.
We shall see below that a similar drastic reduction occurs, when we consider only spatially and temporally averaged functions $J^I=\int dt d^2x \cJ^I_{\pm}$.

Importantly, the two extra currents $\cJ^I_{\pm}$ have a clear meaning in
the CFT. The current $G^I(z)$ reduces by construction to
$\Gamma^I$ times the Green's function $G^{\one}(z_0) $ on the
boundary, and clearly $\bar{G}^I(z)$ is its hermitian conjugate.
The current $\cJ_+^I$ reduces at the boundary to $
\cJ_+^I=\Gamma^IM_{+,0}SM_{+,0}^{-1}$. Thus $\cJ_+^I$ sets the
normalization of the linear system (\ref{eq:12b}). The interesting
current is the current $\cJ_-^I$. Using that
$\bar{S}=\bar{S}^{-1}$,  it can be seen to reduce on the boundary
to the combination $\bar{\cJ}^{\one}_+\bar{G}^{\one}\Gamma^I
G^{\one}$. Thus,
$\left(\bar{\cJ}^{\one}_+\right)^{-1}\cJ_-^{\one}$ is the norm
squared of the Green's function, i.e. the probability density of
the off-shell process.

For an off-shell process or a correlation function the
norm-squared has no real functional meaning. However, we are
specifically interested in solutions in the absence of an external
source, i.e. the {\em on-shell} correlation functions. In that
case the analysis is quite different. The on-shell condition is
equivalent to choosing momenta to saturate the pole in the Green's
function, i.e. it is precisely choosing dual AdS solutions whose
leading external source $A_{\pm}$ vanishes. Then $M_+$ and $M_-$
are no longer independent, but $M_{+,0} = \delta B_+/\delta
\Psi_+^{(hor)}= -\frac{i\mu\gamma^0}{2m+1} M_{-,0}S$. As a
consequence all boundary values of $\cJ_{-}^I(z_0),
G^I(z_0),\bar{G}^I(z_0)$ become proportional; specifically using
$S=\gamma^0$ one has that
\begin{eqnarray}
\cJ^0_-(z_0)|_{on-shell}=\frac{(2m+1)}{\mu} \gamma^0 G^{\one}(z_0)|_{on-shell}
\end{eqnarray}
 is the ``on-shell'' Green's function. Now, the meaning of the  on-shell correlation function is most evident in thermal backgrounds. It equals the density of states $\rho(\ome(k))=-\frac{1}{\pi}{\rm Im}G_R$ times the Fermi-Dirac distribution 
\cite{LL}
\begin{eqnarray}
  \label{eq:27}
\left.
\tr i\gamma^0  G_F^t(\ome_{bare}(k),k) \right|_{\rm{on-shell}}
&=& 2\pi
f_{FD}\left(\frac{\ome_{bare}(k)-\mu}{T}\right)\rho(\ome_{bare}(k))
\end{eqnarray}
For a Fermi liquid with the defining off-shell Green's function
(\ref{zprop}), we have $\ome_{bare}(k_F)-\mu\equiv\ome=0$ and
$\rho(\ome_{bare}(k)) = Z_{z_0}
\delta^2(k-k_F)\delta(\ome)+\ldots$. Thus we see that the boundary
value of $\cJ_-^{(0)}(z_0)|_{on-shell}=Z f_{FD}(0)\del^3(0)$
indeed captures the pole strength, times a product of
distributions. This product of distributions can be absorbed in
setting the normalization. An indication that this is correct is
that the determining equations for $G^I,~\bar{G}^I,~\cJ^I_{\pm}$
remain unchanged if we multiply $G^I,~\bar{G}^I,~\cJ^I_{\pm}$ on
both sides with $M_{+,0}$. If $M_{+,0}$ is unitary it is just a
similarity transformation. However, from the definition of the
Green's function, one can see that this transformation precisely
removes the pole. This ensures that we obtain finite values for
$G^I,~\bar{G}^I,~\cJ^I_{\pm}$ at the specific pole-values
$\ome_F,k_F$ where the distributions would naively blow up.

\subsubsection{Boundary conditions and normalizability}
\label{sec:bound-cond-norm}

We have shown that a normalizable solution to $\cJ_-^{(0)}$ from
the equations (\ref{eq:12b}) correctly captures the pole strength
directly. However, 'normalizable' is still defined in terms of an
absence of a source for the fundamental Dirac field $\Psi_{\pm}$
rather than the composite fields $\cJ^0_{\pm}$ and $G^I$. One would prefer to determine normalizability directly from the boundary behavior of the composite fields. This can be done.  
Under the assumption that the electrostatic potential $\Phi$
is regular,  i.e.
\begin{equation}
\label{eq:4c}
  \Phi=\mu- \rho z+\ldots
  \end{equation}
the composite fields behave near $z=0$ as
\begin{eqnarray}
  \label{eq:3b}
 \cJ^0_{+} &=& j_{3- 2m} z^{3- 2m}+ j_4^{+}z^4 + j_{5+ 2m}z^{5+ 2m}+\ldots,\non
\cJ^0_{-} &=& j_{5- 2m} z^{5- 2m}+ j_4^{-}z^4 + j_{3+2m}z^{3+ 2m}+\ldots,\non
  G^I &=& I_{4-2m}z^{4-2m}+I_3 z^3+ I_{4+2m}z^{4+2m} +I_{5}z^{5}+\ldots,
\end{eqnarray}
with the identification
\begin{eqnarray}
  \label{eq:4b}
j_{3-2m} = |A_+|^2,~ &j_4^+ = A_+^\dagger B_++B_+^\dagger
A_+,~&j_{5+2m} = |B_+|^2~, \non j_{3+2m} = |A_-|^2,~ &j_4^-=
A_-^\dagger B_-+B_-^\dagger A_-,~ &j_{5-2m} = |B_-|^2~, \non
I_{4-2m} = \bar{A}_+ A_-+ \bar{A}_- A_+,~ &I_3 = \bar{A}_+ B_-+
\bar{B}_- A_+,~ &I_{4+2m} = \bar{B}_+ B_-+\bar{B}_- B_+,~ \non
 &&I_5 = \bar{B}_+ A_-+\bar{A}_-
 B_+~.
\end{eqnarray}
A 'normalizable' solution in $\cJ_-^{(0)}$ is thus defined by the
vanishing of {\em both} the leading {\em and} the subleading term.

\section{An AdS Black hole with Dirac Hair}
\setcounter{equation}{0}

Having determined a set of AdS evolution equations and boundary
conditions that compute the pole strength $Z$ directly through the
currents $\cJ_-^{(0)}(z)$ and $G^I(z)$, we can now try to construct the AdS dual of a system with
finite fermion density, including backreaction. As we remarked in the beginning of section \ref{sec:dynam-dirac-field}, the demand that the solutions be normalizable means that the
construction of the AdS black hole solution with a finite single
fermion wavefunction is analogous to the construction of a
holographic superconductor \cite{Hartnoll:2008vx} with the role of
the scalar field now taken by the Dirac field. The starting point therefore is the
charged AdS${}_4$ black-hole background \eqref{metric} and we should show that at low temperatures this AdS Reissner-Nordstr\"{o}m
black hole is unstable towards a solution with a finite Dirac
profile. We shall do so in a simplified ``large charge'' limit where we
ignore the gravitational dynamics, but as is well known from holographic superconductor studies (see e.g. \cite{Hartnoll:2008vx,Hartnoll:2008kx,Hartnoll:2009sz}) this limit already captures much of the essential physics. In a companion article
\cite{ours3} we will construct the full backreacted groundstate
including the gravitational dynamics.

In this large charge non-gravitational limit
the equations of motion for the action \eqref{eq:2a} reduce to those of $U(1)$-electrodynamics coupled to a
fermion with charge $g$ in the background of this black hole:
\begin{eqnarray}
  \label{eq:2bb}
  D_{M}F^{MN} &=& ige_A^N\bar{\Psi}\Gamma^A\Psi~, \non
0&=&e_A^M\Gamma^A(D_M+i{\rm e}gA_M)\Psi+m\Psi~.
\end{eqnarray}
Thus the vielbein $e_A^M$ and and covariant derivative $D_M$
remain those of the fixed charged AdS black hole metric
(\ref{metric}), but the vector-potential now contains a background
piece $A_0^{(bg)}$ plus a first-order piece
$A_M=A_M^{(bg)}+A^{(1)}_M$, which captures the effect of the charge carried by the fermions.   

Following our argument set out in previous section that it is more convenient to work with the currents $\cJ^I_{\pm}(z), G^I(z)$ instead of trying to solve the Dirac equation directly, we shall first rewrite this coupled non-trivial set of equations of motion in terms of the currents while at the same time using symmetries to reduce the complexity. Although a system at finite fermion density need not be
homogeneous, the Fermi liquid ground state is. It therefore natural
to make the ansatz that the final AdS solution is static and
preserves translation and rotation along the boundary. As the
Dirac field transforms non-trivially under rotations and boosts,
we cannot make this ansatz in the strictest sense. However, in
some average sense which we will make precise, the solution should
be static and translationally invariant. 
Then translational and
rotational invariance allow us to set $A_i=0$, $A_z=0$, whose
equations of motions will turn into contraints for the remaining
degrees of freedom. Again denoting $A_0=\Phi$,
the equations reduce to the following after the projection onto $\Psi_{\pm}=\hlf(1\pm\Gamma^{Z})\Psi$.
\begin{eqnarray}
\label{de}
\pa_z^2\Phi &=& \frac{-gL^3\alp}{z^3\sqrt{f}}\left(\bar{\Psi}_+i\gamma^0\Psi_++\bar{\Psi}_-i\gamma^0\Psi_-\right) ~,\non
\left(\partial_z+\mathcal{A}_\pm\right)\Psi_\pm&=&\mp\slT\Psi_\mp
\end{eqnarray}
with
\begin{eqnarray}
\label{not1b}\mathcal{A}_\pm&=&-\frac{1}{2z}\left(3-\frac{zf^\prime}{2f}\right)\pm\frac{mL}{z\sqrt{f}}~, \non
\label{not2b}\slT&=&\frac{i(-\ome +g\Phi)}{\alpha f}\gamma^0+\frac{i}{\alp\sqrt{f}}k_i\gamma^i~.
\end{eqnarray}
as before.

The difficult part is to ``impose'' staticity and rotational invariance
 for the non-invariant spinor. This can be done by rephrasing the dynamics in terms of fermion
 current bilinears, rather than the fermions themselves. We shall first do so rather heuristically, and then show that the equations obtained this way are in fact the flow equations for the Green's functions and composites $\cJ^I(z),~G^I(z)$ constructed in the previous section.
In terms of the local vector currents\footnote{In our conventions $\bar{\Psi} =
\Psi^{\dag}i\gamma^0$.}
\begin{equation}
\label{teqn}J^\mu_{+}(x,z)=\bar{\Psi}_+(x,z)i\gamma^\mu\Psi_+(x,z)~,~~J^\mu_{-}(x,z)= \bar{\Psi}_-(x,z)i\gamma^\mu\Psi_-(x,z)~,
\end{equation}
or equivalently
\begin{equation}
\label{teqn2}J^\mu_{+}(p,z)=\int d^3 k\bar{\Psi}_+(-k,z)i\gamma^\mu\Psi_+(p+k,z)~,~~J^\mu_{-}(p,z)=\int d^3k \bar{\Psi}_-(-k,z)i\gamma^\mu\Psi_-(p+k,z)~.
\end{equation}
rotational invariance means
that spatial components $J^i_{\pm}$ should vanish on the solution
--- this solves the constraint from the $A_i$ equation of
motion, and the equations can be rewritten in terms of $J^0_{\pm}$
only. 
Staticity and rotational invariance in addition demand that the bilinear momentum $p_{\mu}$ vanish. In other words, we are only considering temporally and spatially averaged densities: $J^{\mu}_{\pm}(z)= \int dt d^2x  \bar{\Psi}(t,x,z)i\gamma^{\mu} \Psi(t,x,z)$. When acting with the Dirac operator on the currents to obtain an effective equation of motion, this averaging over
the relative frequencies $\ome$ and momenta $k_i$ also sets all terms with explicit
$k_i$-dependence to zero.\footnote{ To see this consider
\begin{eqnarray}
  \label{eq:9}
  (\pa+2\cA_{\pm})\Psi_{\pm}^\dagger(-k)\Psi_{\pm}(k)=\mp \frac{\Phi}{f} \left(\Psi_{-}^{\dagger}i\gamma^0\Psi_{+} + \Psi_{+}^{\dagger}i\gamma^0\Psi_{-}\right)+\frac{ik_i}{\sqrt{f}}\left( \Psi_-^{\dagger}\gamma^i\Psi_{+}-\Psi_+^{\dagger}\gamma^i\Psi_{-}\right)~.
\end{eqnarray}
The term proportional to $\Phi$ is relevant for the solution. The
dynamics of the term proportional to $k_i$ is
\begin{eqnarray}
  \label{eq:11}
  (\pa+\cA_++\cA_-) (\Psi^\dagger_-\gamma^i\Psi_+-\Psi^\dagger_+\gamma^i\Psi_-)= - 2i \frac{k^i}{\sqrt{f}}(\Psi_+^\dagger\gamma^0\Psi_++\Psi_-^\dagger\gamma^0\Psi_-) ~.
\end{eqnarray}
The integral of the RHS over $k^i$ vanishes by the assumption of
translational and rotational invariance. Therefore the LHS of
(\ref{eq:11}) and thus the second term in eq. (\ref{eq:9}) does so
as well. } Restricting to such averaged currents and absorbing a factor of $g/\alpha$ in
$\Phi$ and a factor of $g\sqrt{L^3}$ in $\Psi_{\pm}$, we
obtain effective equations of motion for the bilinears directly
\begin{eqnarray}
\label{jpmeqn}\left(\partial_z+2\mathcal{A}_\pm\right)J_\pm^0&=&\mp\frac{\Phi}{f}I~,\non
\label{ieq}\left(\partial_z+\mathcal{A}_++\mathcal{A}_-\right)I&=&\frac{2\Phi}{f}(J_+^0-J_-^0)~,\non
\pa_z^2 \Phi &=& -\frac{1}{z^3\sqrt{f}}(J_+^0+J_-^0)~,
\end{eqnarray}
with $I = \bar{\Psi}_-\Psi_{+}+\bar{\Psi}_+\Psi_-$, and all fields
are real. The remaining constraint from the $A_z$ equation of motion decouples.
It demands Im$(\bar{\Psi}_+\Psi_-)=\frac{i}{2}(\bar{\Psi}_-\Psi_+-\bar{\Psi}_+\Psi_-)=0$.
What the equations \eqref{ieq} tell us is that for nonzero $J^0_{\pm}$ there is a charged
electrostatic source for the vector potential $\Phi$ in
the bulk.

Before we motivate the effective equations \eqref{ieq} at a more fundamental level, we note that these equations contain more information than just current conservation $\pa_{\mu}J^{\mu}=0$. In an isotropic and static background, current conservation is trivially satisfied since $\pa_{\mu}J^{\mu}= \pa_0J^0 = -i\int d\ome e^{-i\ome t}\ome J^0(\ome)=0$ as $J^0(\ome\neq0)=0$. Also, note that we have scaled out the electromagnetic coupling. 
AdS${}_4$/CFT${}_3$ duals for which the underlying string theory
is known generically have $g=\kap/L$ with $\kap$ the gravitational
coupling constant as defined in (\ref{eq:2a}). Thus, using
standard AdS${}_4$/CFT${}_3$ scaling, a finite charge in the new
units translates to a macroscopic original charge of order $L/\kap
\propto N^{1/3}$. This large charge demands that backreaction of the fermions in terms of its bilinear is taken into account as a source for $\Phi$.

The justification of using \eqref{ieq} to construct the AdS dual of a regular Fermi liquid is the connection between local fermion bilinears and the CFT Green's function. The complicated flow equations \eqref{eq:12b} reduce precisely to the first two equations in \eqref{ieq} upon performing the spacetime averaging and the trace, i.e. $J_\pm^0 = \int d^3k {\rm Tr} \cJ^0_{\pm}$ and $I = \int d^3k {\rm Tr} \left(G^{\one}+\bar{G}^{\one}\right)$. Combined with the demand that we only consider normalizable solutions and the proof that $\cJ_-^0$ is proportional to the pole-strength, the radial evolution equations \eqref{ieq} are the (complicated) AdS recasting of the RG-flow for the pole-strength.
This novel interpretation ought to dispel some of the a priori worries about our unconventional treatment of the fermions through their semi-classical bilinears. There is also support from the gravity side, however. Recall that for conventional many-body systems and fermions in particular one first populates a certain set of states and then tries to compute the macroscopic properties of the collective. In a certain sense the equations \eqref{ieq} formulate the same program but in opposite order: one computes the generic wavefunction charge density with and by imposing the right boundary conditions, i.e normalizability, one selects only the correct set of states. This follows directly from the
equivalence between normalizable AdS modes and quasiparticle poles
that are characterized by well defined distinct momenta $k_F$ (for
$\ome=\ome_F\equiv 0$). The demand that any non-trivial Dirac hair
black hole is constructed from normalizable solutions of the
composite operators (i.e. their leading and subleading asymptotes
vanish\footnote{One can verify that the discussion in section \ref{sec:bound-cond-norm} holds also for fully backreacted solutions. The derivation there builds on the assumption that the boundary behavior of
the electrostatic potential is regular. It
is straightforward to check in \eqref{ieq} that indeed precisely for normalizable solutions, i.e.  in the absence
of explicit fermion-sources, when { both} the leading and
subleading terms in $J^0_{\pm}$ and $I$ vanish, the boundary
behavior the scalar potential remains regular,
as required.}) thus means that one imposes a superselection rule on the
spatial averaging in the definition of $J_{\pm}^I$:
\begin{eqnarray}
  \label{eq:16}
  J_{\pm}^0(z)|_{normalizable} &\equiv& \int d^3k \bar{\Psi}_{\pm}(-k)i\gamma^0\Psi_{\pm}(k)|_{normalizable} \non
&=& \int d^3k \,\delta^2(|k|-|k_F|) |B_{\pm}^{(0)}(k)|^2 z^{4+2m\pm1} +\ldots
\end{eqnarray}
We see that the constraint of normalizability  from the bulk point
of the view, implies that one selects precisely the on-shell bulk
fermion modes as the building blocks of the density $J_{\pm}^0$.

In turn this means that the true system that eqs. (\ref{ieq})
describe is somewhat obscured by the spatial averaging. Clearly
even a single fermion wavefunction is in truth the full set of
two-dimensional wavefunctions whose momentum $k^i$ has length
$k_F$. However, the averaging could just as well be counting more,
as long as there is another set of normalizable states once the
isotropic momentum surface $|k|=|k_F|$ is filled. Pushing this
thought to the extreme, one could even speculate that the system
(\ref{ieq}) gives the correct quantum-mechanical description of
the many-body Fermi system: the system which gravitational reasoning
suggests is the true groundstate of the charged AdS black hole in
the presence of fermions.

To remind us of the ambiguity introduced by spatial
averaging, we shall give the boundary coefficient of normalizable
solution for $J_-^0=\int d^3k \cJ^0_-$ a separate name. The
quantity $ \cJ_-^0(z_0)$ is proportional to the pole strength,
which via Migdal's relation quantifies the characteristic
occupation number discontinuity at $\ome_F\equiv 0$. We shall
therefore call the coefficient $\int d^3k |B_-|^2|_{normalizable}
= \Delta n_F$.

\subsubsection{Thermodynamics}

At a very qualitative level the identification
$J_-^0|_{norm}(z)\equiv \Delta n_F z^{3+2m} +\ldots$ can be argued
to follow from thermodynamics as well. From the free energy for an
AdS dual solution to a Fermi liquid, one finds that the charge
density directly due to the fermions is
\begin{eqnarray}
  \label{eq:12}
  \rho_{total} = -2\frac{\pa}{\pa \mu} F &=& \frac{-3}{2m+1}\frac{
\Delta n_F }{z_0^{-1-2m}} +\rho +\ldots ~,
\end{eqnarray}
with $z_0^{-1}$ the UV-cutoff as before. The cut-off dependence is
a consequence of the fact that the system is interacting, and one
cannot truly separate out the fermions as free particles. Were one
to substitute the naive free fermion scaling dimension
$\Delta=m+3/2=1$, the cutoff dependence would vanish and the
identification would be exact.

We can thus state that in the interacting system there is a
contribution to the charge density from a finite number of
fermions proportional to
\begin{eqnarray}
\rho_F= \frac{-3}{2\Delta-2}\frac{\Delta n_F}{z_0^{2-2\Delta}}+\ldots~,
\end{eqnarray}
although this contribution formally vanishes in the limit where we send the UV-cutoff $z_0^{-1}$ to infinity.

To derive eq. (\ref{eq:12}), recall that the free energy is equal to
minus the on-shell action of the AdS dual theory. Since we
disregard the gravitational backreaction, the Einstein term in the
AdS theory will not contain any relevant information and we
consider the Maxwell and Dirac term only. We write the action as
\begin{eqnarray}
  \label{eq:6}
  S= \int_{z_0}^1 \sqrt{-g}\left[\frac{1}{2}A_{N}D_MF^{MN}-\bar{\Psi}\slD\Psi-m\bar{\Psi}\Psi  \right] + \oint_{z=z_0} \sqrt{-h} \left(\bar{\Psi}_+\Psi_- + \frac{1}{2}
A_{\mu}n_{\alp}F^{\alp\mu}\right)~,
\end{eqnarray}
where we have included an explicit fermionic boundary term that
follows from the AdS/CFT dictionary \cite{ours} and $n_{\alp}$ is
a normal vector to the boundary. The boundary action is not
manifestly real, but its on-shell value which contributes to the
free energy is real. Recall that the imaginary part of
$\bar{\Psi}_+\Psi_-$ decouples from eqs. (\ref{ieq}). The boundary
Dirac term in (\ref{eq:6}) is therefore equal to $I=2 {\rm
Re}(\bar{\Psi}_+\Psi_-)$.

To write the free energy  in terms of the quantities $\mu,~\rho$
and $\Delta n_F$, note that the on-shell bulk Dirac action
vanishes. Importantly the bulk Maxwell action does contribute to
the free energy. Its contribution is
\begin{eqnarray}
  \label{eq:10}
  F_{bulk} &=& \lim_{z_0\rar 0}\int^1_{z_0} dzd^3x\left[ \frac{1}{2} \Phi \pa_{zz} \Phi \right]_{on-shell} \non
&=& - \lim_{z_0\rar 0}\int^1_{z_0} dzd^3x\left[ \frac{1}{2z^3\sqrt{f}} \Phi (J_+^0+J_-^0) \right]_{on-shell}~,
\end{eqnarray}
where we have used the equation of motion (\ref{ieq}). This
contribution should be expected, since the free energy should be
dominated by infrared, i.e. near horizon physics. Due to the
logarithmic singularity in the electrostatic potential (Eq.
(\ref{horbnd}) this bulk contribution diverges, but this
divergence should be
 compensated by gravitational backreaction.
At the same time the singularity is so mild, however, that the
free energy, the integral of the Maxwell term, remains finite in
the absence of the Einstein contribution.

Formally, i.e. in the limit $z_0\rar 0$, the full free energy
arises from this bulk contribution~(\ref{eq:10}). The relation
(\ref{eq:12}) between the charge density and $\Delta n_F$ follows
only from the regularized free energy, and is therefore only a
qualitative guideline. Empirically, as we will show, it is
however, a very good one (see Fig \ref{freeE} in the next
section). Splitting the regularized bulk integral in two
\begin{eqnarray}
  \label{eq:13}
  F_{bulk} = \int_{z^{*}}^1 dzd^3x\left[ \frac{1}{2z^3\sqrt{f}} \Phi (J_+^0+J_-^0) \right]_{on-shell}+ \lim_{z_0\rar 0} \int_{z_0}^{z_*} dzd^3x\left[ \frac{1}{2z^3\sqrt{f}} \Phi (J_+^0+J_-^0) \right]_{on-shell}~,
\end{eqnarray}
we substitute the normalizable boundary behavior of
$\Psi_+=B_+ z^{5/2+ m}+\ldots$, $\Psi_-=B_- z^{3/2+m}+\ldots$ and $\Phi=\mu-\rho z+\ldots $, and obtain for the regularized free energy
\begin{eqnarray}
  \label{eq:7b}
  F=F_{horizon}(z_*)+ \lim_{z_0\rar 0}\int_{z_0}^{z_{*}} d^3x dz \left[\frac{-1}{2z^{3}} \mu|B_-|^2z^{3+2m}+\ldots\right] + \oint \frac{d^3x}{z_0^3}\left[ - \bar{B}_+B_- z_0^{4+2m} + \frac{1}{2} \mu\rho z^3_0\right]~.
\end{eqnarray}
Using that $B_+=- i\mu \gamma^0 B_-/(2m+1)$ (eq. (\ref{eq:5})),
the second bulk term and boundary contribution are proportional,
and the free energy schematically equals
\begin{eqnarray}
  \label{eq:7}
  F=F^{horizon}+ \lim_{z_0\rar 0} \int d^3x\left[ \frac{3\mu}{2(2m+1)}\bar{B}_-i\gamma^0B_- z_0^{1+2m} - \frac{1}{2} \mu\rho\right]~.
\end{eqnarray}
With the UV-regulator $z_0^{-1}$ finite, this yields the charge
density in Eq. (\ref{eq:12}) after one recalls that
$\bar{B}_-=B^\dagger_-i\gamma^0$. 

With the derived rule that the AdS dual to a Fermi liquid has a
nonzero normalizable component in the current $J_-^0$, we will now
construct an AdS solution that has this property: an AdS black hole with
Dirac hair. Ignoring backreaction, these are solutions to the
density equations \eqref{ieq}. In its simplest form the
interpretation is that of the backreaction due to a single fermion
wavefunction, but as explained the spatial averaging of the
density combined with the selection rule of normalizability could
be capturing a more general solution.

\subsection{At the horizon: Entropy collapse to a Lifshitz solution}
\label{sec:entr-coll-lifsh}

Before we can proceed with the construction of non-trivial Dirac
hair solutions to Eqs. \eqref{ieq}, we must consider the boundary
conditions at the horizon necessary to solve the system.
Insisting that the right-hand-side of the dynamical equations
\eqref{ieq} is subleading at the horizon, the near-horizon
behavior of $J_\pm^0,~I,~\Phi$ is:
\begin{eqnarray}
J_\pm^0&=&J_{hor,\pm} (1-z)^{-1/2}+\ldots,\non
I &=& I_{hor}
(1-z)^{-1/2}+\ldots,\non
\Phi &=&
\Phi_{hor}^{(1)}(1-z)\ln(1-z)+(\Phi_{hor}^{(2)}-\Phi^{(1)}_{hor})(1-z)+\ldots\label{horbnd}.
\end{eqnarray}
If we insist that $\Phi$ is regular at the horizon $z=1$, i.e. $
\Phi_{hor}^{(1)}=0$, so that the electric field is finite, the
leading term in $J^0_{\pm}$ must vanish as well, i.e.
$J_{hor,\pm}=0$, and the system reduces to a free Maxwell field in
the presence of an AdS black hole and there is no fermion density
profile in the bulk. Thus in order to achieve a nonzero fermion
profile in the bulk, we must have an explicit source for the
electric-field on the horizon. Strictly speaking, this invalidates
our neglect of backreaction as the electric field and its energy
density at the location of the source will be infinite. As we
argued above, this backreaction is in fact expected to resolve the
finite ground-state entropy problem associated with the presence
of a horizon. The backreaction should remove the horizon
completely, and the background should resemble the horizonless
metrics found in \cite{polchinski,Goldstein:2009cv,electronstar};
the same horizon logarithmic behavior in the electrostatic
potential was noted there. Nevertheless, as the divergence in the
electric field only increases logarithmically as we approach the
horizon, and our results shall hinge on the properties of the
equations at the opposite end, near the boundary, we shall continue
to ignore it here. We shall take the sensibility of our result
after the fact as proof that the logarithmic divergence at the
horizon is indeed mild enough to be ignored.

The identification of the boundary value of $J_-^0$ with the Fermi
liquid characteristic occupation number jump  $\Delta n_F$ rested
on the insistence that the currents are built out of AdS Dirac
fields. This deconstruction also determines a relation between the
horizon boundary conditions of the composite fields
$J^0_{\pm},~I$. If $\Psi_{\pm}(z)= C_{\pm}(1-z)^{-1/4}+ \ldots$
then $J_{hor,\pm}=C_{\pm}^2$ and $I_{hor}=C_+C_-$. As the solution
$\Phi^{(1)}_{hor}$ is independent of the solution
$\Phi^{(2)}_{hor}$ which is regular at the horizon, we match the
latter to the vector-potential of the charged AdS black hole:
$\Phi^{(2)}_{hor}=-2gq \equiv g\mu_0/\alp$. Recalling that
$\Phi^{(1)}_{hor} =-(J_{hor,+}+J_{hor,-})$, we  see that the
three-parameter family of solutions at the horizon in terms of
$C_{\pm}$, $\Phi^{(2)}_{hor}$ corresponds to the three-parameter
space of boundary values $A_+,~B_-$ and $\mu$ encoding a
fermion-source, the fermion-response/expectation value and the
chemical potential.


We can now search whether within this three-parameter family a
finite normalizable fermion density solution with vanishing source
$A_+=0$ exists for a given temperature $T$ of the black hole.

\subsection{A BH with Dirac hair}\label{seccond}

The equations are readily solved numerically with a
shooting method from the horizon. We consider both an uncharged
AdS-Schwarzschild solution and the charged AdS
Reissner-Nordstr\"{o}m solution. Studies of bosonic condensates in
AdS/CFT without backreaction have mostly been done in the AdS-Schwarzschild (AdSS)
background (\cite{Hartnoll:2008vx, Hartnoll:2008kx} and references
therein). An exception is \cite{aljohn}, which also considers the
charged RN black hole. As is explained in \cite{aljohn}, they
correspond to two different limits of the exact solution: the AdSS
case requires that $\Delta n_F \gtrsim \mu$ that is, the total charge
of the matter fields should be dominant compared to the charge
 of the black hole. On the other hand, the RN limit is
appropriate if $\Delta n_F\ll \mu $. It ignores the effect of the
energy density of the charged matter sector on the charged black
hole geometry. The AdS Schwarzschild background is only reliable
near $T_c$, as at low temperatures the finite charged
fermion density is comparable to $\mu$. The RN case is under
better control for low temperatures, because near $T=0$ the
chemical potential can be tuned to stay larger than fermion
density.

We shall therefore focus primarily on the solution
in the background of an AdS RN black hole, i.e. the system with a
heat bath with chemical potential $\mu$ --- non-linearly
determined by the value of $\Phi^{(2)}_{hor}=\mu_0$ at the horizon
--- which for low $T/\mu$ should show the characteristic $\Delta
n_F$ of a Fermi liquid. The limit in which we may confidently
ignore backreaction is
 $\Phi^{(1)}_{hor}\ll \mu_0$ for $T\lesssim \mu_0$ --- for AdSS the appropriate limit is $\Phi^{(1)}_{hor}\ll T$ for $\mu_0 \ll T$.

\subsubsection{Finite fermion density solutions in AdS-RN}\label{seccondrn}

\begin{figure}[t]
(A)
\hspace*{-.25in}
\raisebox{.15in}{
\includegraphics[width=3.5in]{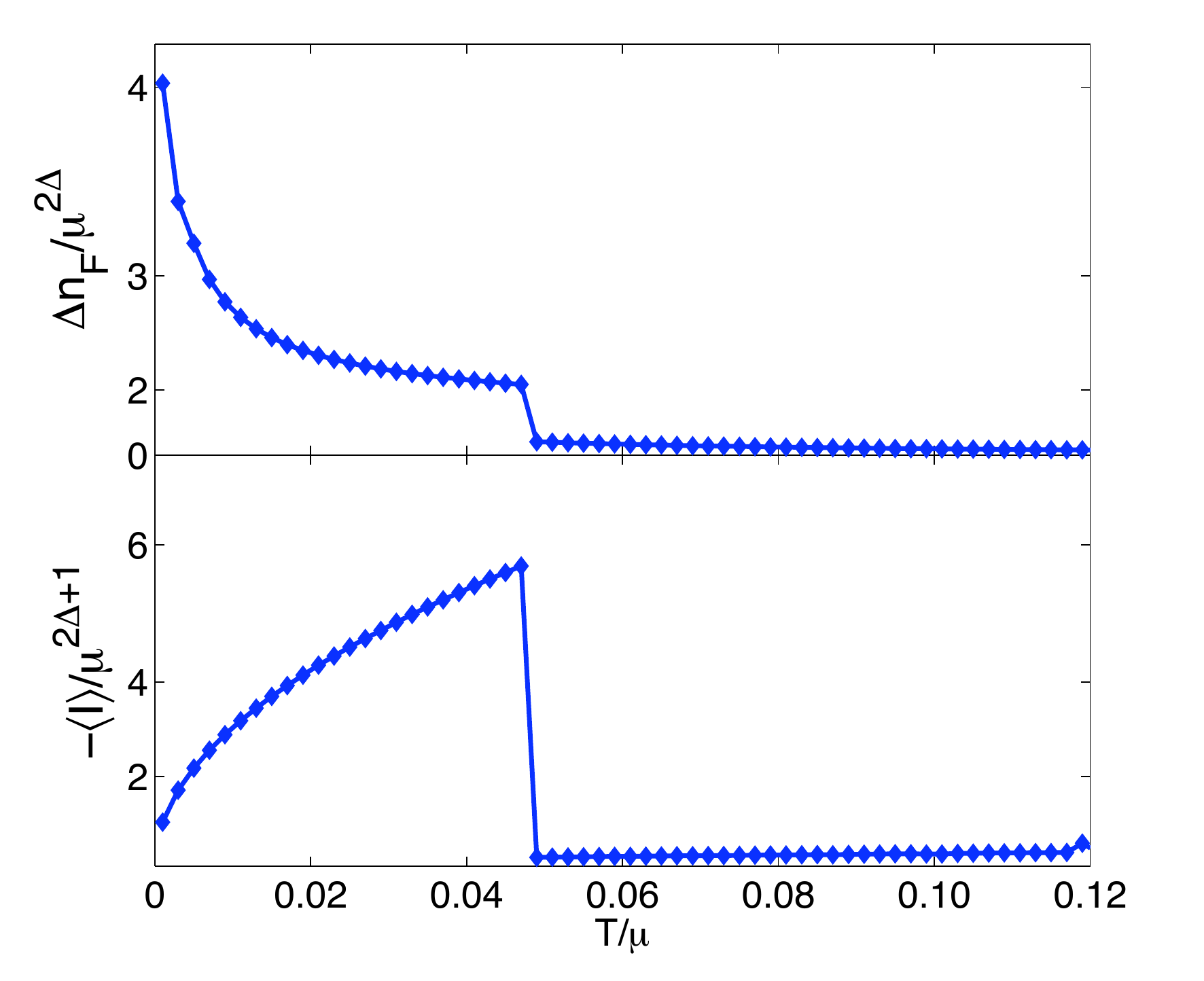}}
(B)
\includegraphics[width=3.25in]{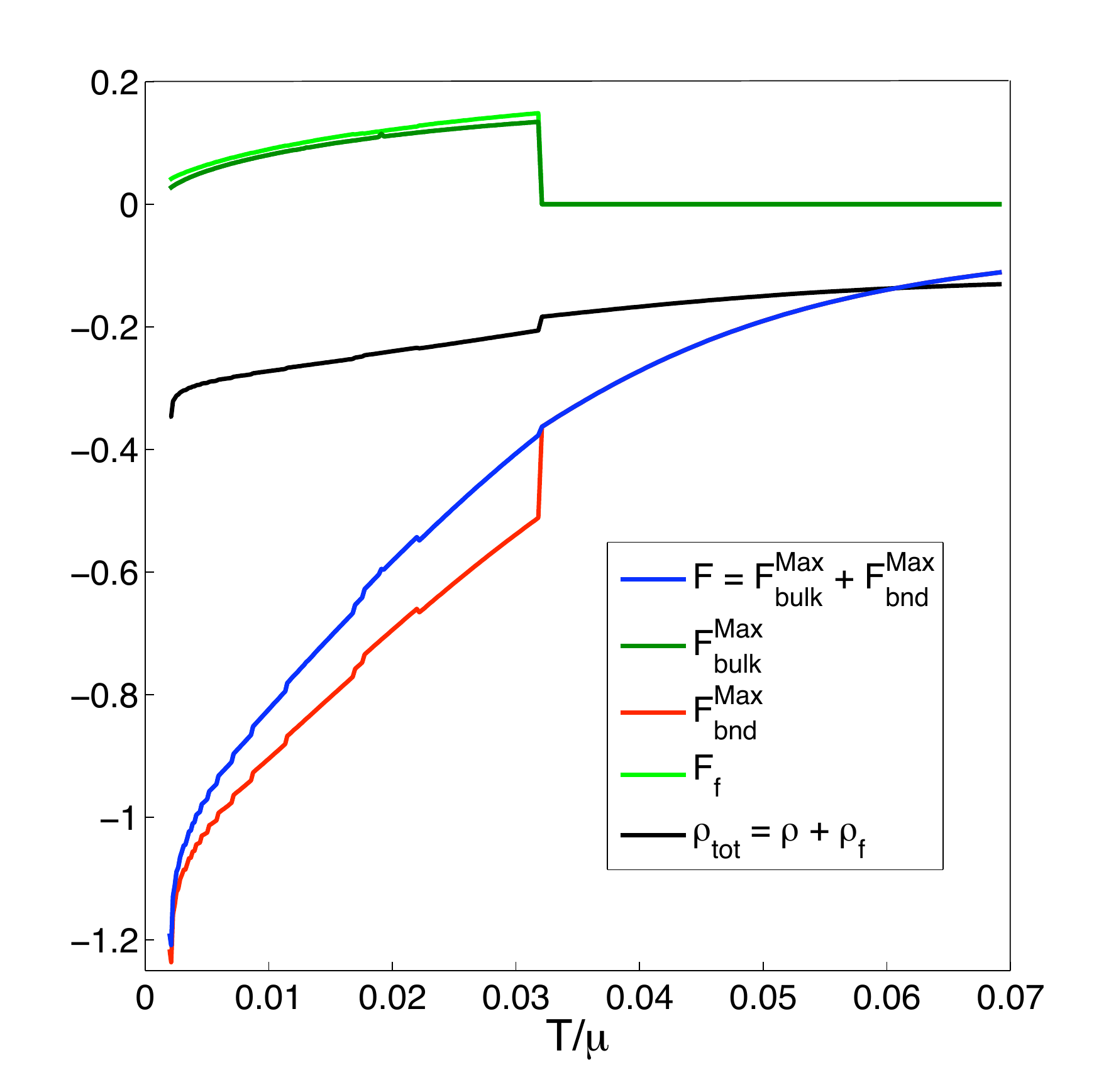}
\caption{\small\label{cond1} (A) Temperature dependence of the
Fermi liquid occupation number discontinuity $\Delta n_F$ and
operator $I$ for a fermionic field of mass $m=-1/4$ dual to an
operator of dimension $\Delta=5/4$. We see a large density for
$T/\mu$ small and discontinuously drop to zero at $T\approx
0.05\mu$. At this same temperature, the proxy free energy
contribution per particle (the negative of $I$) vanishes. (B)
\label{freeE} The free energy $F=F^{fermion}+F^{Maxwell} $
(Eq.~(\ref{eq:6})) as a function of $T/\mu$ ignoring the
contribution from the gravitational sector. The blue curve shows
the total free energy $F=F^{Maxwell}$, which is the sum of a bulk
and a boundary term.  The explicit fermion contribution
$F_{fermion}$ vanishes, but the effect of a non-zero fermion
density is directly encoded in a non-zero $F^{Maxwell}_{bulk}$.
The figure also shows this bulk $F^{Maxwell}_{bulk}$ and the
boundary contribution $F^{Maxwell}_{bulk}$ separately and how they
sum to a continuous $F_{total}$. Although  formally the explicit
fermion contribution $F_{f}\sim I$ in equation (\ref{eq:7})
vanishes, the bulk Maxwell contribution is captured remarkably
well by its value when the cut-off is kept finite. The light-green
curve in the figure shows $F_f$ for a finite $z_0 \sim 10^{-6}$.
For completeness we also show the total charge density, Eq.
(\ref{eq:12}). The dimension of the fermionic operator used in
this figure is $\Delta=1.1$.}
\end{figure}

\begin{figure}[ht]
\begin{center}
\includegraphics[width=2.5in]{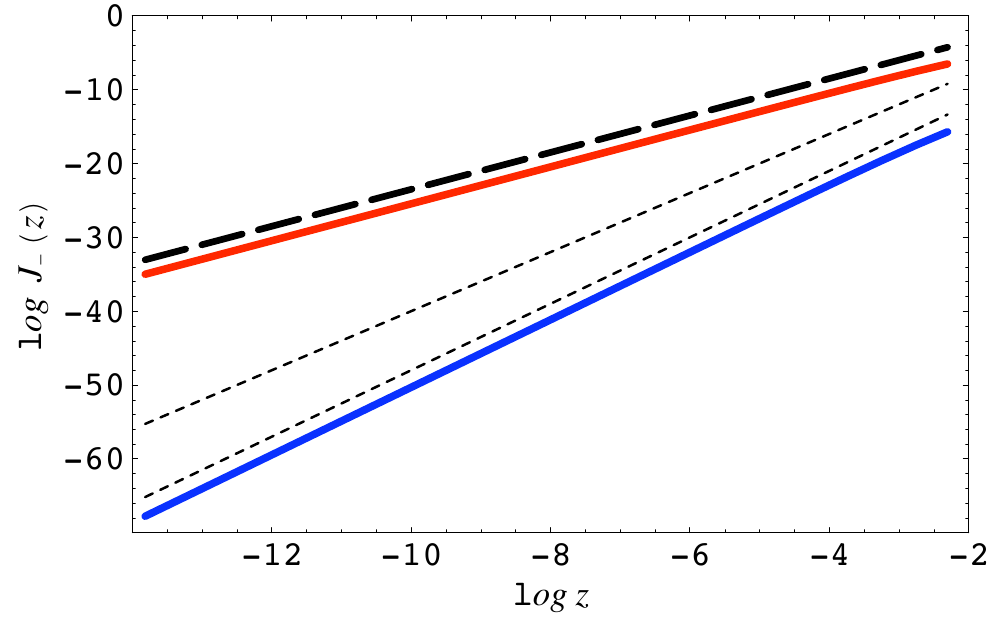}
\end{center}
\caption{\small\label{diff} The boundary behaviour of $J_-(0)$ in
for a generic solution (blue) to Eqs. \eqref{ieq} and a
normalizable Dirac-hair solution (red) for $m=-1/4$ in the
background of an AdS-RN black hole with $\mu/T=128.8$. The dotted
lines show the scaling $z^{11/2}$ and $z^{4}$ of the leading and
subleading terms in an expansion of $J_-^0(z)$ near $z=0$; the
dashed line shows the scaling $z^{5/2}$ of the subsubleading
expansion whose coefficient is $|B_-(\ome_F,k_F)|^2$. That the
Dirac hair solution (red) scales as the subsubleading solution
indicates that the current $J_-^0$ faithfully captures the density
of the underlying normalizable Dirac field. }
\end{figure}

Fig. \ref{cond1} shows the behavior of the occupation number
discontinuity $n_F \equiv  |B_-|^2$ and the fermion free-energy
contribution $I$ as a function of temperature in a search for
normalizable solutions to Eqs \eqref{ieq} with the aforementioned
boundary conditions. We clearly see a first order transition to a
finite fermion density, as expected. The underlying Dirac field
dynamics can be recognized in that the normalizable solution for
$J_-^0(z)$ which has no leading component near the boundary by
construction, also has its subleading component vanishing (Fig.
\ref{diff}).\footnote{Although the Dirac hair solution has charged
matter in the bulk, there is no Higgs effect for the bulk gauge
field, and thus there is no direct spontaneous symmetry breaking
in the boundary. Indeed one would not expect it for the Fermi
liquid groundstate. There will be indirect effect on the
conductivity similar to \cite{electronstar}. We thank Andy
O'Bannon for his persistent inquiries to this point.}

\begin{figure}[ht]
(A)\raisebox{.01in}{\includegraphics[width=3in]{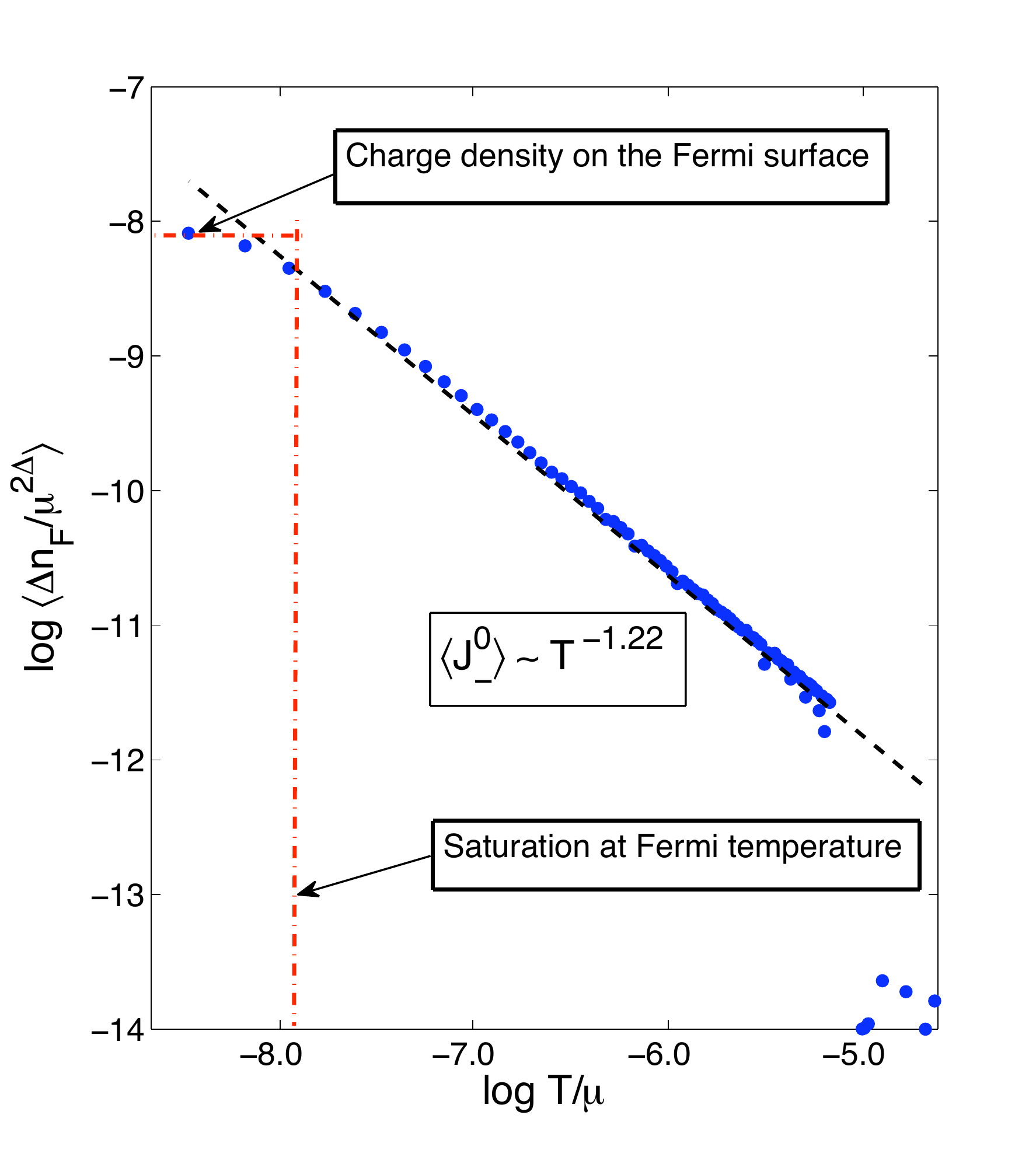}} (B)
\hspace*{-.2in}
\raisebox{.15in}{\includegraphics[width=3.5in]{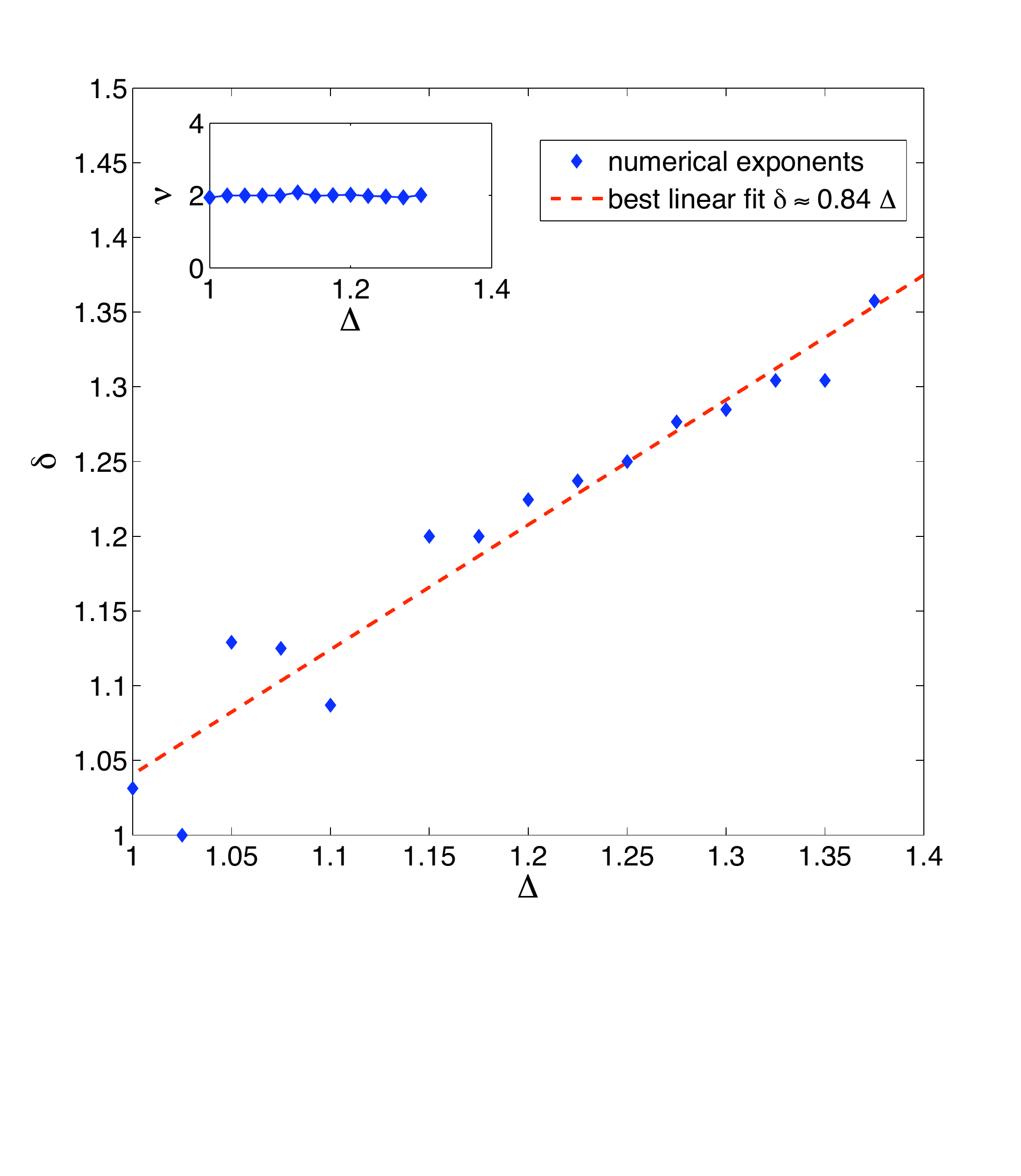}}
\caption{\small\label{deld}
 (A) Approximate power-law scaling of the Fermi liquid characteristic occupation number discontinuity  $\Delta n_F/\mu^{2\Delta} \sim T^{-\delta}$ as a function of $T/{\mu}$
for $\Delta=5/4$. This figure clearly shows the saturation of the
density at very low $T/\mu$. The saturation effect is naturally
interpreted as the influence of the characteristic Fermi energy.
(B) The scaling exponent $\delta$ for different values of the
conformal dimension $\Delta$. There is a clear correlation, but
the precise relation cannot be determined  numerically. The
scaling exponent of the current $I/\mu^{2\Delta+1}\sim T^{-1/\nu}$
obeys $\nu=2$ with great accuracy, on the other hand (Inset).}
\end{figure}

Analyzing the transition in more detail in Fig. \ref{deld}, we
find:
\begin{enumerate}
\item[1.] The dimensionless number discontinuity $\Delta n_F/\mu^{2\Delta}$ scales as
$T^{-\delta}$ in a certain temperature range $T_F<T<T_c$, with
$\delta>0$ depending on $g$ and $\Delta$, and $T_F$ typically very
small. At $T=T_c>T_F$ it drops to zero discontinuously,
characteristic of a first order phase
transition.

\item[2.] At low temperatures, $0<T<T_F$, the power-law growth
comes to a halt and ends with a plateau where
$\Delta n_F/\mu^{2\Delta}\sim\mathrm{const.}$ (Fig. \ref{deld}A). It is
natural to interpret this temperature as the Fermi temperature of
the boundary Fermi liquid.

\item[3.] The fermion free energy contribution $I/\mu^{2\Delta+1}$ scales as $T^{1/\nu}$ with
$\nu>1$ for $0<T<T_c$, and drops to zero discontinuously at $T_c$.
As $I$ empirically equals minus the free
energy per particle, it is natural that $I(T=0)=0$, and
this in turn supports the identification of $\Delta n_F(T=0)$ as the
step in number density at the Fermi energy.
\end{enumerate}

One expects that the exponents $\delta,\nu$ are controlled by the
conformal dimension $\Delta$.\footnote{The charge $g$ of the
underlying conformal fermionic operator scales out of the
solution.} The dependence of the exponent $\delta$ on the
conformal dimension is shown in Fig. \ref{deld}A. While a
correlation clearly exists, the data are not conclusive enough to
determine the relation $\delta=\delta(\Delta)$. The clean power
law $T^{-\delta}$ scaling regime is actually somewhat puzzling.
These values of the temperature, $T_F<T<T_c$, correspond to a
crossover between the true Fermi liquid regime for $T<T_F$ and the
conformal phase for $T>T_c$, hence there is no clear ground for a
universal scaling relation for $\delta$, which seems to be
corroborated by the data (Fig. \ref{deld}B). At the same time, the
scaling exponent $\nu$ appears to obey $\nu=2$ with great
precision (Fig. \ref{deld}B, inset) independent of $\Delta$ and
$g$.

\begin{figure}[t]
(A)
\hspace*{-.6in}
\raisebox{1.6in}{
\begin{tabular}{c}
\includegraphics[width=2in]{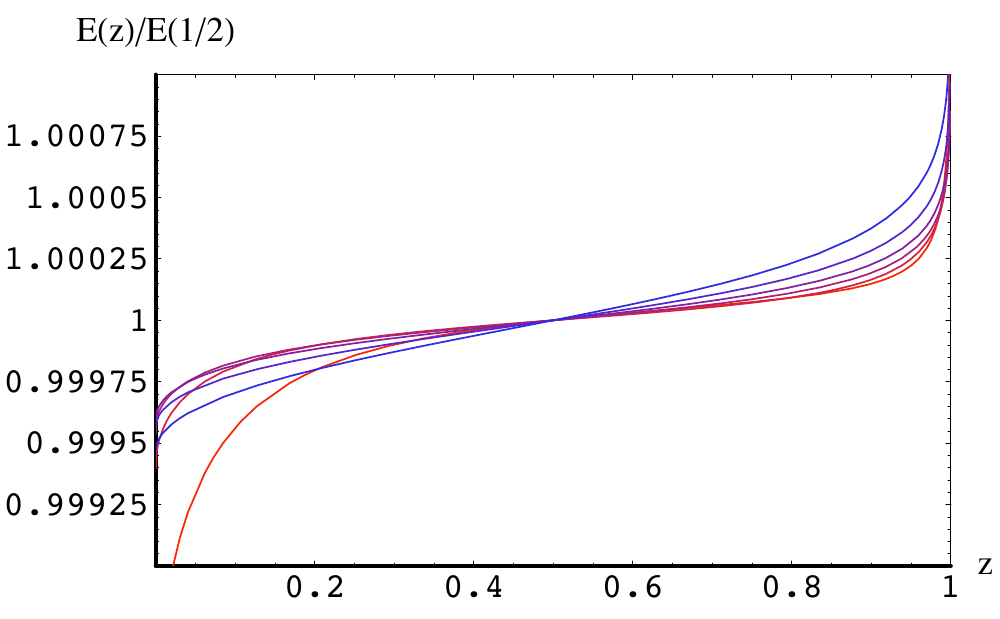}\\[.2in]
\includegraphics[width=2in]{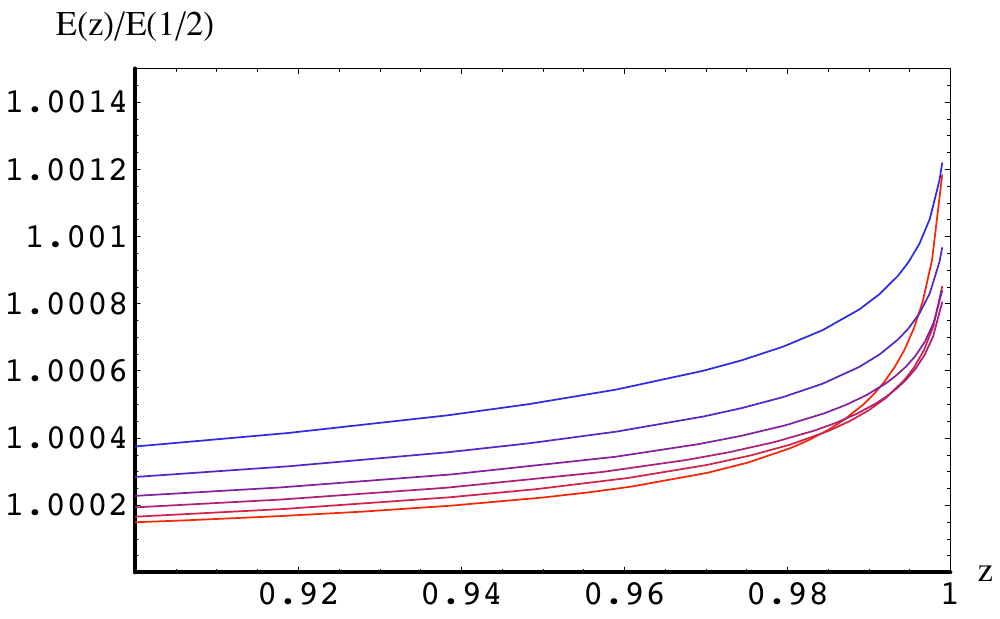}
\end{tabular}}
(B)
\hspace*{-0.6in}
\raisebox{.15in}{
\includegraphics[width=3in]{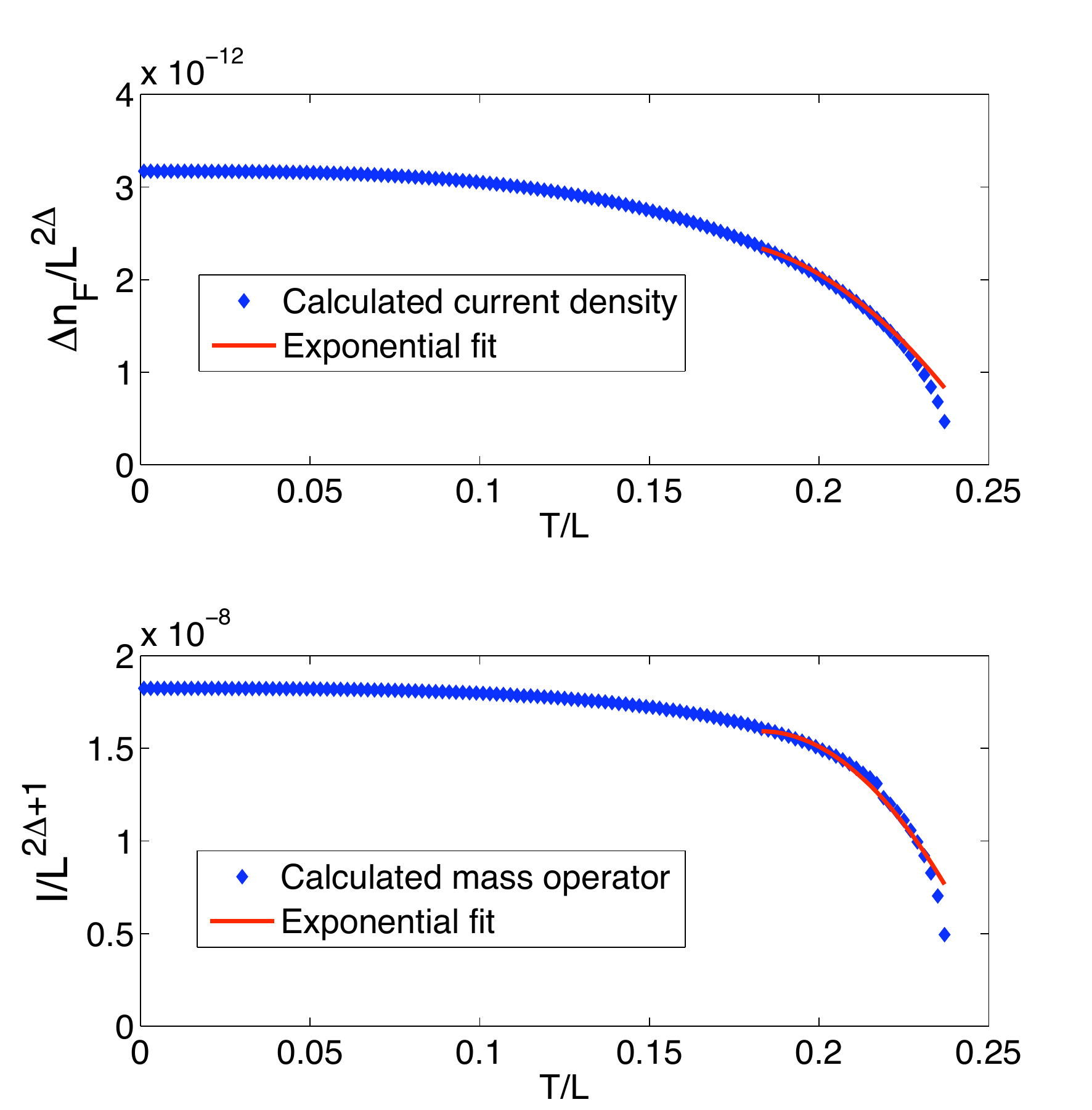}}
\hspace*{-.2in}
(C)
\hspace*{-.4in}
\raisebox{1.6in}{
\begin{tabular}{c}
\includegraphics[width=2in]{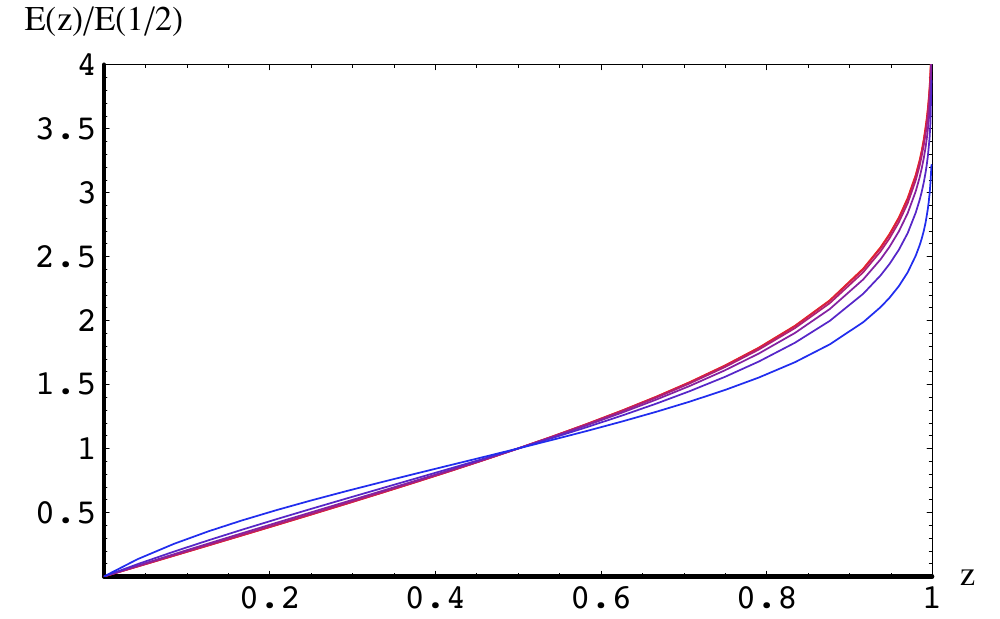}\\[.2in]
\includegraphics[width=2in]{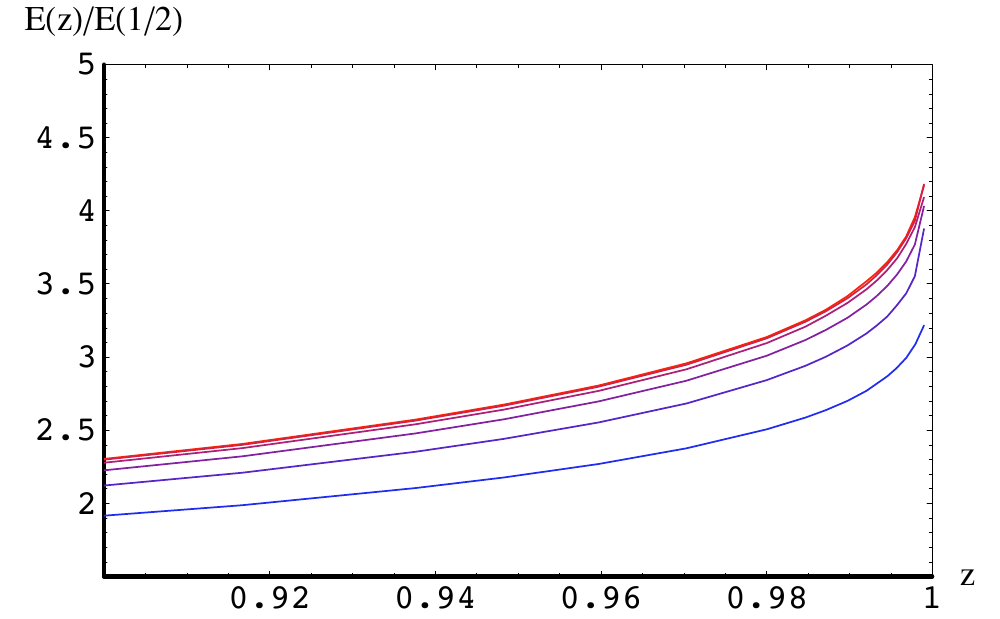}
\end{tabular}}
\caption{\small(A) \label{logsinga} The radial electric field $-E_z=\partial
\Phi/\partial z$, normalized to the midpoint value
$E_z(z)/E_z(1/2)$ for whole interior of the finite fermion density AdS-RN solution (upper) and near the horizon (lower). One clearly sees the soft, log-singularity at the horizon. The colors correspond to increasing temperatures from
$T=0.04\mu$ (lighter) to $T=0.18\mu$ (darker), all with $\Delta=1.1$.
(B)\label{adsscond} The occupation number jump $\Delta n_F$ and free energy contribution
 $I$ as a function of temperature in AdS-Schwarzschild. We see the
jump $\Delta n_F$ saturate at low temperatures and fall off at high $T$.
An exponential fit to the data (red curve) shows that in
the critical region the fall-off is stronger than exponential, indicating that the transition is first order.
The conformal dimension of the fermionic operator is $\Delta=1.1$. (C) \label{adssbulk} The radial electric field
$-E_z=\partial \Phi/\partial z$, normalized to the midpoint value
($E_z(z)/E_z(1/2)$) for the finite fermion density AdS-Schwarzschild background. The divergence of
the electric field $E_z$ is again only noticeable near the horizon and can be
neglected in most of the bulk region.}
\end{figure}

A final consideration, needed to verify the existence of a finite
fermion density AdS solution dual to a Fermi liquid, is to show
that the ignored backreaction stays small. In particular, the
divergence of the electric field at the horizon should not affect
the result. The total bulk electric field $E_z=-\partial_z\Phi$ is
shown in Fig. \ref{logsinga}A, normalized by its value at $z=1/2$.
The logarithmic singularity at the horizon is clearly visible. At
the same time, the contribution to the  total electric field from
the charged fermions is negligible even very close to the
horizon.\footnote{It is of the order $10^{-4}$, starting from
$z=0.9999$. We have run our numerics using values between
$1-10^{-6}$ and $1-10^{-2}$ and found no detectable difference in
quantities at the boundary.} This suggests that our results are
robust with respect to the details of the IR divergence of the
electric field.

The diverging backreaction at the horizon is in fact the gravity
interpretation of the first order transition at $T_c$: an
arbitrarily small non-zero density leads to an abrupt change in
the on shell bulk action. As the latter is the free energy in the
CFT, it must reflect the discontinuity of a first order
transition. A full account of the singular behavior at the horizon
requires self-consistent treatment including the Einstein
equations. At this level, we can conclude that the divergent
energy density at the horizon implies that the near-horizon
physics becomes substantially different from the AdS${}_2$ limit
of the RN metric. It is natural to guess that the RN horizon
disappears completely, corresponding to a ground state with zero
entropy, as hypothesized in \cite{polchinski}. This matches the
expectation that the finite fermi-density solution in the bulk
describes the Fermi-liquid. The underlying assumption in the above
reasoning is that the total charge is conserved.

\subsubsection{Finite fermion density in AdSS}

For completeness, we will describe the finite fermion-density
solutions in the AdS Schwarzschild geometry as well. In these
solutions the charge density is set by the density of fermions
alone. They are therefore not reliable at very low temperatures
$T\ll T_c$ when gravitational backreaction becomes important. The
purpose of this section is to show the existence of finite density
solutions does not depend on the presence of a charged black-hole
set by the horizon value $\Phi^{(2)}_{hor} =\mu_0$, but that the
transition to a finite fermion density can be driven by the
charged fermions themselves.

Fig. \ref{adsscond}B shows the nearly instantaneous development of
a non-vanishing expectation value for the occupation number
discontinuity $\Delta n_F$ and $I$ in the AdS Schwarzschild
background. The rise is not as sharp as in the RN background. It
is, however, steeper than exponential, and we may conclude that
the system undergoes a discontinuous first order transition to a
AdS Dirac hair solution. The constant limit reached by the fermion
density as $T\to 0$ has no meaning as we cannot trust the solution
far away from $T_c$.

The backreaction due to the electric field divergence at the
horizon can be neglected, for the same reason as before (Fig.
\ref{adssbulk}C).

\subsection{Confirmation from fermion spectral functions}\label{secspec}

If, as we surmised, the finite fermion density phase is the true
Fermi-liquid-like ground state, the change in the fermion spectral
functions should be minimal as the characteristic quasi-particle
peaks are already present in the probe limit, i.e. pure AdS
Reissner-Nordstr\"{o}m \cite{Liu:2009dm, ours}. Fig.
\ref{back250fig} shows that quasiparticle poles near $\omega=0$
with similar analytic properties can be identified in both the
probe limit, i.e. pure AdS-RN case and the AdS-RN Dirac-hair solution.
The
explanation for this similarity is that the electrostatic
potential $\Phi$  almost completely determines the spectrum, and
the change in $\Phi$ due to the presence of a finite fermion
density is quite small. Still, one expects that the finite fermion
density system is a more favorable state. This indeed follows from
a detailed comparison between the spectral functions $A(\omega;k)$
in the probe limit and the fermion-liquid phase (Fig.
\ref{back250fig}). We see that:

\begin{figure}[t]
\mbox{\hspace*{2.3in}\includegraphics[width=5in]{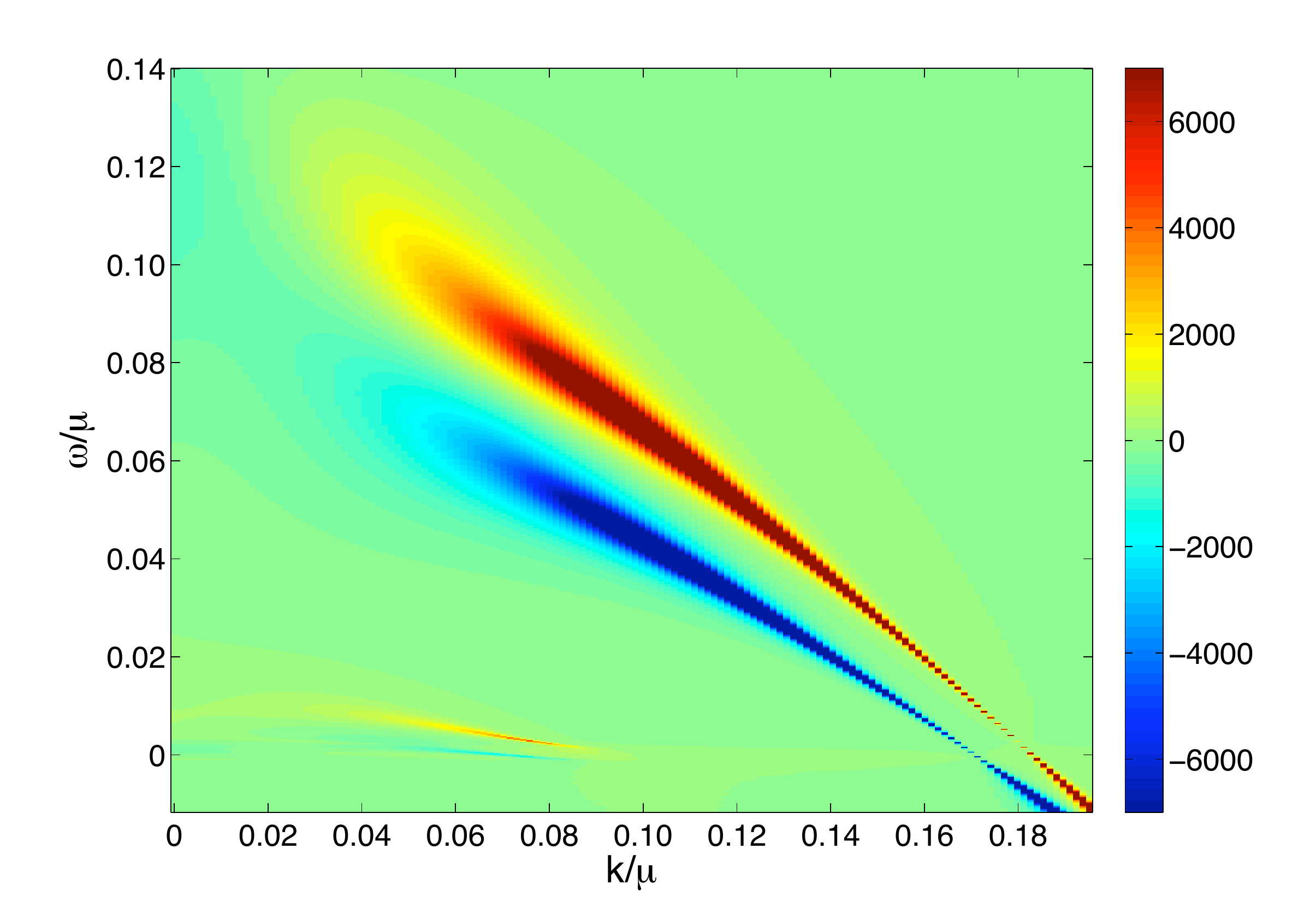}
\hspace*{-7.5in}\includegraphics[width=3.5in]{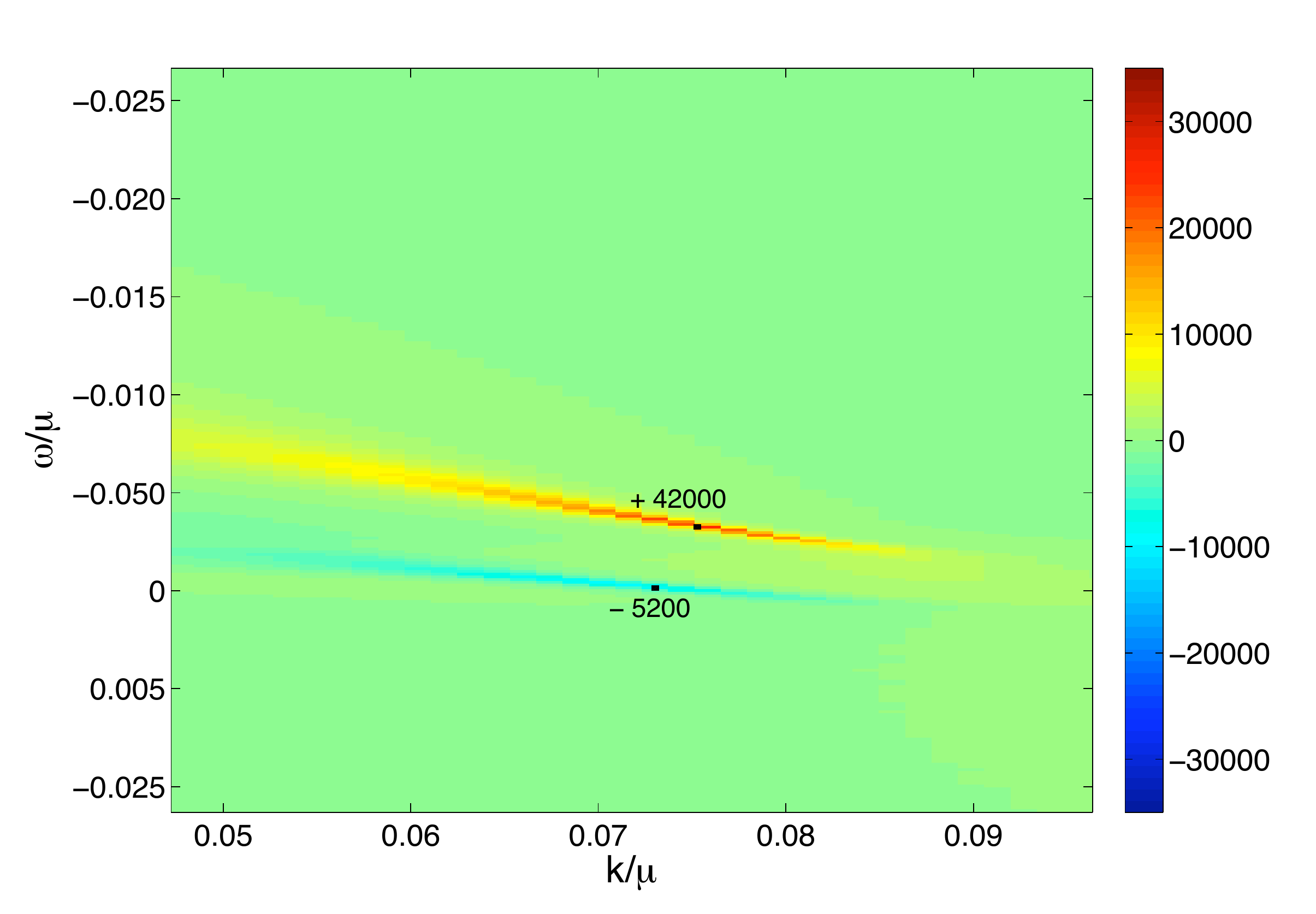}}
\caption{\small\label{back250fig} The single-fermion spectral
function in the probe limit of pure AdS Reissner-Nordstr\"{o}m
(red/yellow) {\em minus} the spectrum in the finite density system
(blue). The conformal dimension is $\Delta=5/4$, the probe charge
$g=2$, and $\mu/T=135$. We can see two quasiparticle poles near
$\omega=0$, a non-FL pole with $k_F^{probe}\simeq 0.11\mu$ and
$k_F^{\Delta n_F}\simeq 0.08\mu$ respectively  and a FL-pole with
$k_F^{probe}\simeq 0.18\mu$ and $k_F^{\Delta n_F}\simeq 0.17\mu$.
The dispersion of both poles is visibly similar between the probe
and the finite density backgroudnd. At the same time, the non-FL
pole has about $8$ times less weight in the finite density
background, whereas the FL-pole has gained about $6.5$ times more
weight.}
\end{figure}

\begin{figure}[t]
(A)
\includegraphics[width=3.25in]{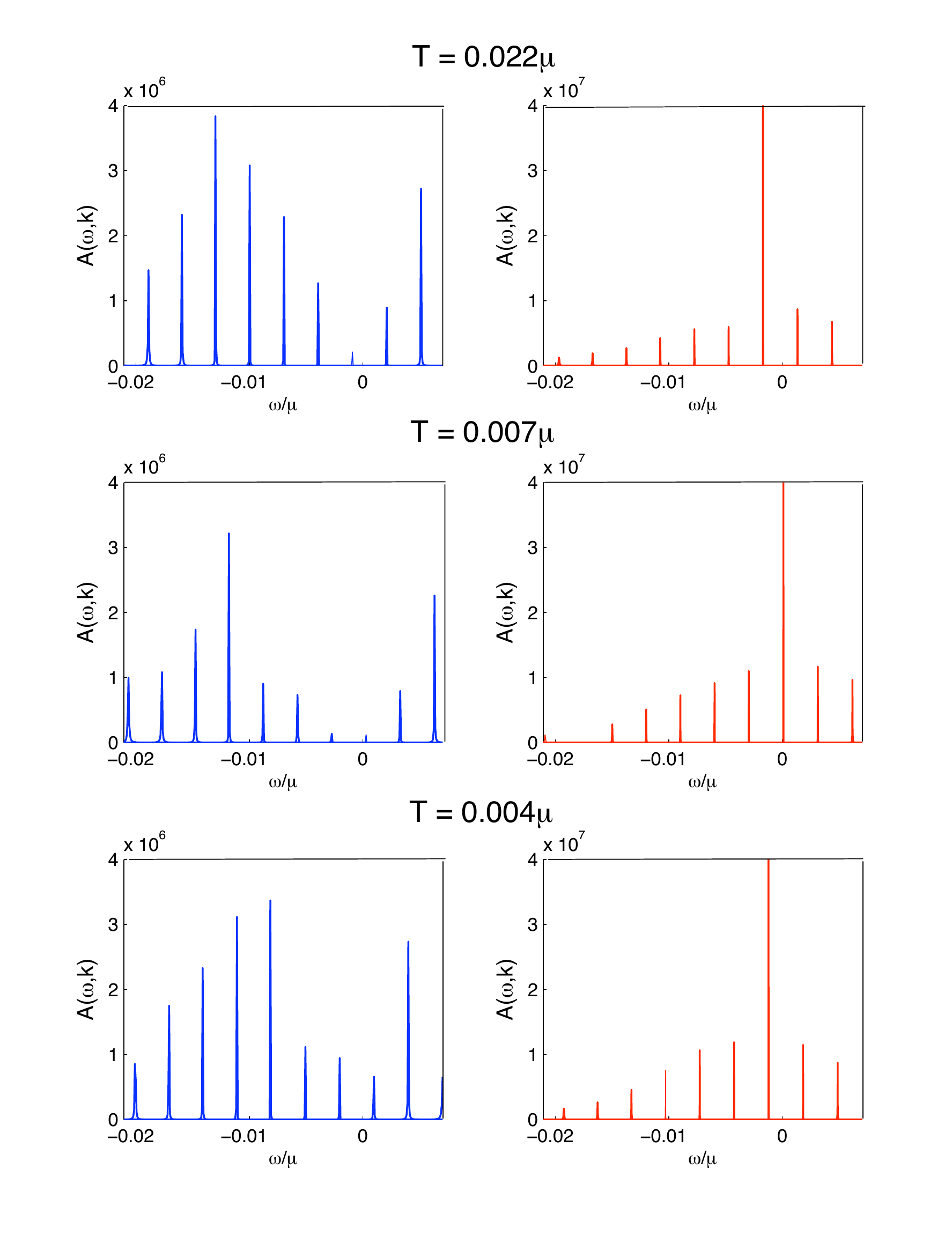}
(B)
\includegraphics[width=3.25in]{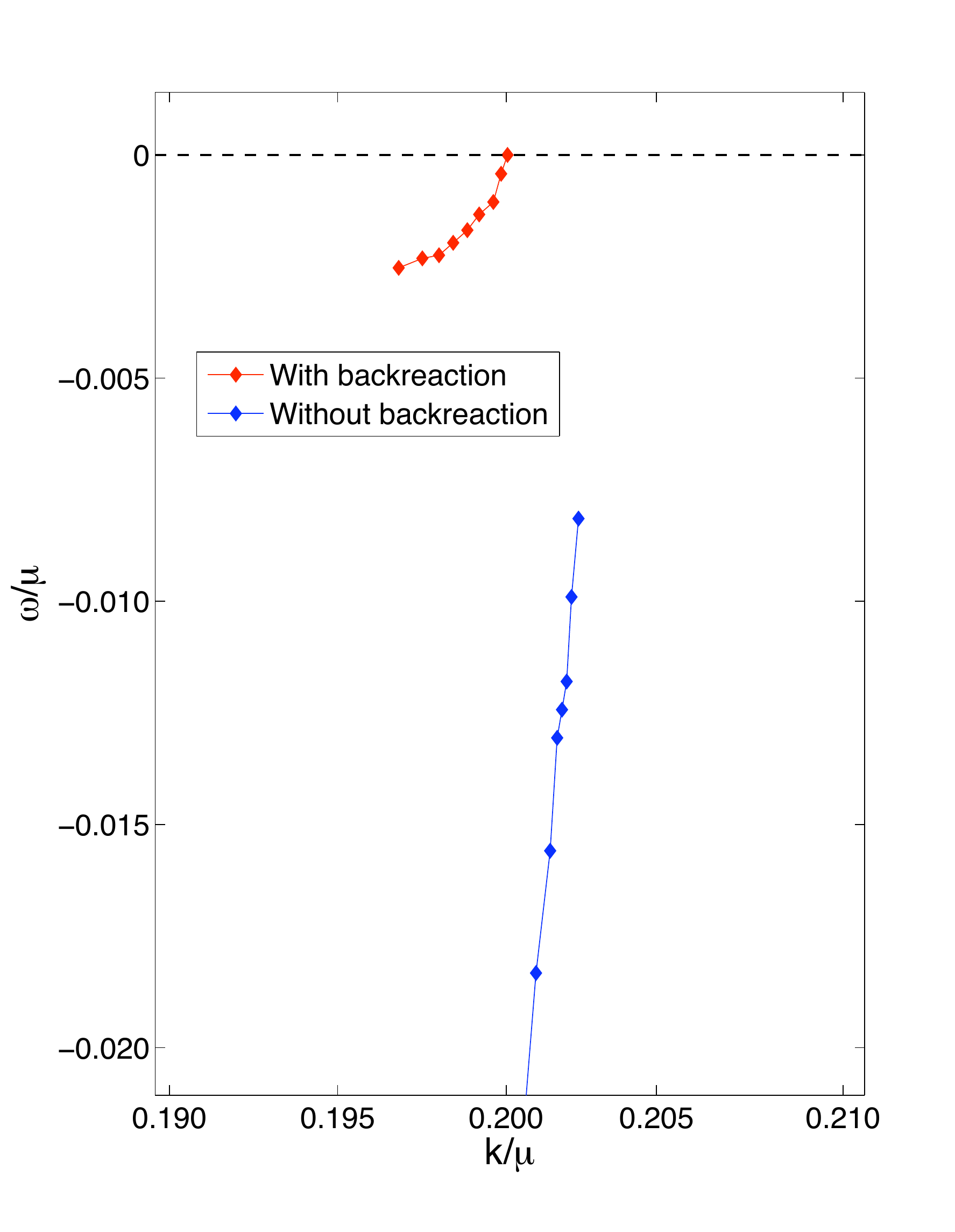}
\caption{\small\label{somefigure} (A) Single fermion spectral
functions near $\omega=0$ in pure AdS Reissner-Nordstr\"{o}m
(blue) and in the finite fermion density background (red). In the
former the position of the maximum approaches $\ome=0$ as $T$ is
lowered whereas in the latter the position of the maximum stays
close to $T=0$ for all values of $T$. (B) Position of the maximum
of the quasiparticle peak in $k$-$\omega$ plane, for different
temperatures and $\Delta=5/4$. The probe limit around a AdS-RN
black hole (blue) carries a strong temperature dependence of the
$\omega_{max}$ value, with $\omega_{max, T\neq 0}\neq 0$. In the
finite fermion density background, the position of the maximum
(red) is nearly independent of temperature and stays at
$\omega=0$.}
\end{figure}

\begin{enumerate}
\item [1.] All quasiparticle poles present in the probe limit are
also present in the Dirac hair phase, at a slightly shifted value
of $k_F$. This shift is a consequence of the change in the bulk
electrostatic potential $\Phi$ due to the presence of the
charged matter. For a Fermi-liquid-like quasiparticle corresponding to
the second pole in the operator with $\Delta=5/4$ and $g=2$ we
find \mbox{$k_F^{probe}-k_F^{\Delta n_F}=0.07\mu$.} The non-Fermi-liquid
pole, i.e. the first pole for the same conformal operator, has
$k_F^{probe}-k_F^{\Delta n_F}=0.03\mu$.

\item [2.] The dispersion exponents $\nu$ defined through
$(\omega-E_F)^2\sim(k-k_F)^{2/\nu}$, also maintain roughly the
same values as both solutions. This is visually evident in the
near similar slopes of the ridges in Fig. \ref{back250fig}. In the
AdS Reissner-Nordstr\"{o}m background, the dispersion coefficients
are known analytically as a function of the Fermi momentum:
$\nu_{k_F}=\sqrt{2\frac{k_F^2}{\mu^2}-\frac{1}{3}+\frac{1}{6}\left(\Delta-3/2\right)^2}$
\cite{Faulkner:2009wj}. The Fermi-liquid-like quasiparticle
corresponding to the second pole in the operator with $\Delta=5/4$
and $g=2$ has $\nu_{k_F}^{probe}=1.02$ vs. $\nu^{\Delta n_F}=1.01$. The
non-Fermi-liquid pole corresponding to the first pole for the same
conformal operator, has $\nu_{k_F}^{probe}\approx 0.10$, and
$\nu^{\Delta n_F}=0.12$.

\item [3.] The most distinct property of the finite density phase
is the redistributed spectral weight of the poles. The non-Fermi
liquid pole reaches its maximum height about $10^4$, an order of
magnitude less than in the probe limit, whereas the second, Fermi
liquid-like pole, increases by an order of magnitude. This
suggests that the finite density state corresponds to the
Fermi-liquid like state, rather than a non-Fermi liquid.

\item [4.] As we mentioned in the introduction,  part of  the
reason to suspect the existence of an AdS-RN Dirac-hair solution
is that a detailed study of spectral functions in AdS-RN reveals
that the quasiparticle peak is anomalously sensitive to changes in
$T$. This anomalous temperature dependence {disappears} in the
finite density solution. Specifically in pure AdS-RN the position
$\ome_{max}$ where the peak height is maximum, denoted $E_F$ in
\cite{ours}, does not agree with the value $\ome_{pole}$, where
the pole touches the real axis in the complex $\ome$-plane, for
any finite value of $T$, and is exponentially sensitive to changes
in $T$ (Fig \ref{somefigure}). In the AdS-RN Dirac hair solution
the location $\ome_{max}$ and the location $\ome_{pole}$ do become
the same. Fig. \ref{somefigure}B shows that the maximum of the
quasiparticle peak always sits at $\omega\simeq 0$ in finite
density Dirac hair solution, while it only reaches this as $T\to
0$ in the probe AdS-RN case.

\end{enumerate}

\section{Discussion and Conclusion}

Empirically we know that the Fermi liquid phase of real matter
systems is remarkably robust and generic. This is corroborated by
analyzing effective field theory around the Fermi surface, but as
it assumes the ground state it cannot explain its genericity. If
the Fermi liquid ground state is so robust, this must also be a
feature of the recent holographic approaches to strongly
interacting fermionic systems. Our results here indicate that this
is so: We have used Migdal's relation to construct AdS/CFT rules
for the holographic dual of a Fermi liquid: the characteristic
occupation number discontinuity $\Delta n_F$ is encoded in the
normalizable subsubleading component of the spatially averaged
fermion density $J_-^0(z) \equiv \int d^3k
\bar{\Psi}(\ome=0,-k,z)i\gamma^0\Psi(\ome=0,k,z)$ near the AdS
boundary. This density has its own set of evolution equations,
based on the underlying Dirac field, and insisting on
normalizability automatically selects the on-shell wavefunctions
of the underlying Dirac-field.

The simplest AdS solution that has a non-vanishing expectation
value for the occupation number discontinuity $\Delta n_F$ is that
of a single fermion wavefunction. Using the density approach ---
which through the averaging appears to describe a whole class of
solutions rather than one specific solution --- we have constructed the
limit of this solution where gravitational backreaction is
ignored. At low black hole temperatures this solution with
fermionic ``Dirac hair'' is the preferred ground state. Through an
analysis of the free-energy, we argue that this gravitational
solution with a non-zero fermion profile precisely corresponds to
a system with a finite density of fermions. A spectral analysis
still reveals a zoo of Fermi-surfaces in this ground state, but
there are indications that in the full gravitationally backreacted
solution only a Landau Fermi-liquid type Fermi surface survives.
This follows in part from the relation between the spectral
density and the Fermi momentum of a particular Landau liquid-like
Fermi surface; it also agrees with the prediction from Luttinger's
theorem. Furthermore, the spectral analysis in the finite density
state shows no anomalous temperature dependence present in the
pure charged black-hole single spectral functions. This also
indicates that the finite density state is the true ground state.

The discovery of this state reveals a new essential component in the
study of strongly coupled fermionic systems through gravitational
duals, where one should take into account the expectation values
of fermion bilinears. Technically the construction of the full
gravitationally backreacted solution is a first point that is
needed to complete our finding. This is under current
investigation. The realization, however, that expectation values
of fermion bilinears can be captured in holographic duals and
naturally encode phase separations in strongly coupled fermion
systems should find a large set of applications in the near
future.

\bigskip
\noindent {\bf Acknowledgements:} We thank F. Benini, J. de Boer,
S. Hartnoll, C. Herzog, E. Gubankova, H. Liu, E. Kiritsis, J. McGreevy, A.
O'Bannon, G. Policastro, and J. Sonner for extensive discussions.
We are very grateful to the KITP Santa Barbara for its generous
hospitality and the organizers and participants of the AdS/CMT
Workshop (July 2009). KS thanks the Galileo Galilei Institute for
Theoretical Physics for the hospitality and the INFN for partial
support during the completion of this work. This research was
supported in part by a VIDI Innovative Research Incentive Grant
(K. Schalm) from the Netherlands Organisation for Scientific
Research (NWO), a Spinoza Award (J. Zaanen) from the Netherlands
Organisation for Scientific Research (NWO) and the Dutch
Foundation for Fundamental Research on Matter (FOM).


\begin{thebibliography}{99}


\bibitem{SFL}
C. M. Varma, Z. Nussinov, W. van Saarloos,
{\em Singular Fermi Liquids}
Phys. Rep. {\bf 361}, 267 (2002)
[arXiv:cond-mat/0103393].

\bibitem{SZ}
J.H. She, J. Zaanen,
{\em BCS Superconductivity in Quantum Critical Metals},
Phys. Rev. {\bf B80}, 184518 (2009)
[arXiv:0905.1225 [cond-mat]].

\bibitem{RV}
H. v. L\"{o}hneysen, A. Rosch, M. Vojta, P. W\"{o}lfle
{\em Fermi-liquid instabilities at magnetic quantum phase transitions},
Rev. Mod. Phys. {\bf 79}, 1015 (2007)
[arXiv:cond-mat/0606317 [cond-mat]].


\bibitem{linear}
M.~Gurvitch, A.~T.~Fiory,
{\em Resistivity of $La1.825$$Sr0.175$$CuO4$ and $YBa2$$Cu3$$O7$ to 1100 K: Absence of saturation and its implications},
Phys.\ Rev.\ Lett.\  {\bf 59}, 1337 (1987).



\bibitem{PhillipsChamon}
P.~Phillips, C.~Chamon,
{\em Breakdown of One-Paramater Scaling in Quantum Critical Scenarios for the High-Temperature Copper-oxide Superconductors},
Phys. Rev. Lett. {\bf 95}, 107002 (2005).
[arXiv:cond-mat/0412179]

\bibitem{JZScience}
J. Zaanen,
{\em Quantum critical electron systems: the unchartered sign worlds} ,
Science {\bf 319}, 1205 (2008).

\bibitem{Luttinger}
J.~M.~Luttinger,
  {\em Fermi Surface and Some Simple Equilibrium Properties of a System of
  Interacting Fermions},
  Phys.\ Rev.\  {\bf 119}, 1153 (1960).

\bibitem{arpes1}
J.~C.~Campuzano, M.~R.~Norman,  M.~Randeria,
``Photoemission in the High-T${}_c$ Superconductors'', in
\textit{Handbook of Physics: Physics of Conventional and
Unconventional Superconductors}, edited by K.~H.~Benneman and
J.~B.~Ketterson, (Springer Verlag, 2004); [arXiv:cond-mat/0209476]

\bibitem{arpes2}
A.~Damascelli, Z.~Hussain,  Z.~X.~Shen,
  {\em Angle-resolved photoemission studies of the cuprate superconductors},
  Rev.\ Mod.\ Phys.\  {\bf 75}, 473 (2003); [arXiv:cond-mat/0208504]


\bibitem{arpes3}
X.~J.~Zhou, T.~Cuk, T.~Devereaux, N.~Nagaosa, Z.-X.~Shen,
``Angle-Resolved Photoemission Spectroscopy on Electronic Structure and Electron-Phonon Coupling in Cuprate Superconductors'' in {\em Handbook of High-Temperature Superconductivity: Theory and Experiment}, edited by J. R. Schrieffer, (Springer Verlag, 2007); [arXiv:cond-mat/0604284]


\bibitem{Liu:2009dm}
  H.~Liu, J.~McGreevy, D.~Vegh,
  {\em Non-Fermi liquids from holography},
  arXiv:0903.2477 [hep-th].

\bibitem{ours}
M.~\v{C}ubrovi\'{c}, J.~Zaanen,  K.~Schalm,
  {\em String Theory, Quantum Phase Transitions and the Emergent Fermi-Liquid},
  Science {\bf 325}, 439 (2009)
  [arXiv:0904.1993 [hep-th]].


\bibitem{Faulkner:2009wj}
  T.~Faulkner, H.~Liu, J.~McGreevy,  D.~Vegh,
  {\em Emergent quantum criticality, Fermi surfaces, and AdS2},
  arXiv:0907.2694 [hep-th].

\bibitem{strangemet}
  T.~Faulkner, N.~Iqbal, H.~Liu, J.~McGreevy,  D.~Vegh,
  {\em From black holes to strange metals},
  arXiv:1003.1728 [hep-th],
Science {\bf 329},  1043 (2010).


\bibitem{Lee:2008xf}
  S.~S.~Lee,
  {\em A Non-Fermi Liquid from a Charged Black Hole: A Critical Fermi Ball},
  Phys.\ Rev.\  D {\bf 79}, 086006 (2009)
  [arXiv:0809.3402 [hep-th]].

\bibitem{Rey:2009ag}
S.~J.~Rey,
  {\em String Theory On Thin Semiconductors: Holographic Realization Of Fermi
  Points And Surfaces},
  Prog.\ Theor.\ Phys.\ Suppl.\  {\bf 177}, 128 (2009)
  [arXiv:0911.5295 [hep-th]].



\bibitem{Gubser:2008px}
  S.~S.~Gubser,
  {\em Breaking an Abelian gauge symmetry near a black hole horizon},
  Phys.\ Rev.\  D {\bf 78}, 065034 (2008)
  [arXiv:0801.2977 [hep-th]].


\bibitem{Hartnoll:2008vx}
  S.~A.~Hartnoll, C.~P.~Herzog,  G.~T.~Horowitz,
  {\em Building a Holographic Superconductor},
  Phys.\ Rev.\ Lett.\  {\bf 101}, 031601 (2008)
  [arXiv:0803.3295 [hep-th]].


\bibitem{Hartnoll:2008kx}
  S.~A.~Hartnoll, C.~P.~Herzog,  G.~T.~Horowitz,
  {\em Holographic Superconductors},
  JHEP {\bf 0812}, 015 (2008)
  [arXiv:0810.1563 [hep-th]].

\bibitem{Hartnoll:2009sz}
  S.~A.~Hartnoll,
  {\em Lectures on holographic methods for condensed matter physics},
  Class.\ Quant.\ Grav.\  {\bf 26}, 224002 (2009)
  [arXiv:0903.3246 [hep-th]].


\bibitem{Finster:2002vk}
  F.~Finster, J.~Smoller,  S.~T.~Yau,
  {\em Non-Existence of Black Hole Solutions to Static, Spherically Symmetric
  Einstein-Dirac Systems - a Critical Discussion},
  arXiv:gr-qc/0211043.

\bibitem{Finster:2000vy}
  F.~Finster, J.~Smoller,  S.~T.~Yau,
  {\em Absence of stationary, spherically symmetric black hole solutions for
  Einstein-Dirac-Yang-Mills equations with angular momentum},
  Adv.\ Theor.\ Math.\ Phys.\  {\bf 4}, 1231 (2002)
  [arXiv:gr-qc/0005028].

\bibitem{Finster:2000ps}
  F.~Finster, J.~Smoller,  S.~T.~Yau,
  {\em The interaction of Dirac particles with non-Abelian gauge fields and
  gravity: Bound states},
  Nucl.\ Phys.\  B {\bf 584}, 387 (2000)
  [arXiv:gr-qc/0001067].

\bibitem{Finster:1999mn}
  F.~Finster, J.~Smoller,  S.~T.~Yau,
  {\em Non-existence of black hole solutions for a spherically symmetric,  static
  Einstein-Dirac-Maxwell system},
  Commun.\ Math.\ Phys.\  {\bf 205}, 249 (1999).

\bibitem{polchinski}
S.~A.~Hartnoll, J.~Polchinski, E.~Silverstein,  D.~Tong,
  {\em Towards strange metallic holography},
  JHEP {\bf 1004}, 120 (2010)
  [arXiv:0912.1061 [hep-th]].


\bibitem{electronstar}
 S.~A.~Hartnoll,  A.~Tavanfar,
  {\em Electron stars for holographic metallic criticality},
  arXiv:1008.2828 [hep-th].


\bibitem{verlinde}
  J.~de Boer, K.~Papadodimas,  E.~Verlinde,
  {\em Holographic Neutron Stars},
  JHEP {\bf 1010}, 020 (2010)
  [arXiv:0907.2695 [hep-th]].


\bibitem{verlinde2}
  X.~Arsiwalla, J.~de Boer, K.~Papadodimas and E.~Verlinde,
  {\em Degenerate Stars and Gravitational Collapse in AdS/CFT},
  arXiv:1010.5784 [hep-th].


\bibitem{Jatkar}
  L.~Y.~Hung, D.~P.~Jatkar and A.~Sinha,
  {\em Non-relativistic metrics from back-reacting fermions},
  Class.\ Quant.\ Grav.\  {\bf 28}, 015013 (2011)
  [arXiv:1006.3762 [hep-th]].

\bibitem{spec}
  M.~Cubrovic, Y.~Liu, K.~Schalm, Y.~W.~Sun and J.~Zaanen,
  {\em Spectral probes of the holographic Fermi groundstate: dialing between the
  electron star and AdS Dirac hair},
  arXiv:1106.1798 [hep-th].


\bibitem{sean}
  S.~A.~Hartnoll, D.~M.~Hofman and D.~Vegh,
  {\em Stellar spectroscopy: Fermions and holographic Lifshitz criticality}
  arXiv:1105.3197 [hep-th].

\bibitem{hong}
  N.~Iqbal, H.~Liu and M.~Mezei,
  {\em Semi-local quantum liquids},
  arXiv:1105.4621 [hep-th].

\bibitem{Iqbal:2009fd}
N.~Iqbal,  H.~Liu,
{\em Real-Time Response in AdS/CFT with Application to Spinors},
Fortsch.\ Phys.\  {\bf 57} (2009) 367
[arXiv:0903.2596 [hep-th]].



\bibitem{contino}
  R.~Contino,  A.~Pomarol,
  {\em Holography for fermions},
  JHEP {\bf 0411}, 058 (2004)
  [arXiv:hep-th/0406257].


\bibitem{LL}
E. M. Lifshitz, L. P. Pitaevskii, {\em Statistical physics. Part 2}, Pergamon Press, Oxford, 1980.

\bibitem{ours3}
M.~\v{C}ubrovi\'{c}, J.~Zaanen,  K.~Schalm,
  {\em Stability of the Fermi Liquid from holography}, to appear.

\bibitem{Goldstein:2009cv}
  K.~Goldstein, S.~Kachru, S.~Prakash,  S.~P.~Trivedi,
  {\em Holography of Charged Dilaton Black Holes},
  JHEP {\bf 1008}, 078 (2010)
  [arXiv:0911.3586 [hep-th]].


\bibitem{aljohn} T.~Albash,  C.~V.~Johnson,
{\em A Holographic Superconductor in an External Magnetic Field},
  JHEP {\bf 0809}, 121 (2008)
  [arXiv:0804.3466 [hep-th]].

\bibitem{Cai:2010tr}
  R.~G.~Cai, Z.~Y.~Nie, B.~Wang,  H.~Q.~Zhang,
  {\em Quasinormal Modes of Charged Fermions and Phase Transition of Black
  Holes},
  arXiv:1005.1233 [gr-qc].


\end{thebibliography}
\end{document}